\documentclass{imsart}

\RequirePackage{amsthm,amsmath}
\RequirePackage[numbers]{natbib}
\RequirePackage[colorlinks,citecolor=blue,urlcolor=blue]{hyperref}
\RequirePackage{graphicx}

\startlocaldefs

\RequirePackage{amssymb,bm}
\RequirePackage[mathscr]{eucal}
\DeclareFontEncoding{FMS}{}{}
\DeclareFontSubstitution{FMS}{futm}{m}{n}
\DeclareFontEncoding{FMX}{}{}
\DeclareFontSubstitution{FMX}{futm}{m}{n}
\DeclareSymbolFont{fouriersymbols}{FMS}{futm}{m}{n}
\DeclareSymbolFont{fourierlargesymbols}{FMX}{futm}{m}{n}
\DeclareMathDelimiter{\hsnorm}{\mathord}{fouriersymbols}{152}{fourierlargesymbols}{147}
\newcommand{\hs}{\big\hsnorm}

\newtheorem*{proposition}{Proposition}

\newcommand{\tr}{{\rm tr}}
\newcommand{\ownint}[4]{{\int_{#1}^{#2} \! #3 \, \mathrm{d}#4}}

\newcommand {\X}{\mathcal{H}}

\usepackage{marginnote}

\endlocaldefs

\begin{document}

\begin{frontmatter}

\title{Covariance-based soft clustering of functional data based on the Wasserstein-Procrustes metric}
\runtitle{Covariance-based functional soft clustering}

\begin{aug}

 \author{\fnms{Valentina} \snm{Masarotto}\ead[label=e1]{v.masarotto@math.leideuniv.nl}}
  \address{Mathematisch Instituut\\ Universiteit Leiden\\ Netherlands\\ \printead{e1}\\}

 \and

\author{\fnms{Guido} \snm{Masarotto}\ead[label=e2]{guido.masarotto@unipd.it}}
  \address{Dipartimento di Scienze Statistiche \\Università di Padova\\Italy\\ \printead{e2}\\}

  \runauthor{V.~Masarotto and G.~Masarotto}

\end{aug}

\begin{abstract}
    We consider the problem of clustering functional data according to
    their covariance structure. We contribute a soft clustering methodology 
    based on the Wasserstein-Procrustes distance, where the in-between cluster variability is penalised by a term proportional to the entropy of the partition matrix. In this way, each covariance operator can be partially classified into more than one group. Such soft classification allows for clusters to overlap, and arises naturally in situations where the separation
    between all or some of the clusters is not well-defined. We also discuss how to estimate the number of groups and to test for the presence of any cluster structure. The algorithm is illustrated using simulated and real data. An \texttt{R} implementation is available in the Supplementary materials.
\end{abstract}

\end{frontmatter}

\tableofcontents

\section{Introduction} 
\label{sec:intro}  
Scientific literature is rich in
methods for classifying observations into subsets such that
similar observations are clustered together, and dissimilar ones are
separated. In this work, we focus on classifying functional observations on the base of their covariance structure. This problem occurs naturally in functional data analysis, for instance, when we are given a collection of $N$ groups of functions in a function space $\mathcal{H}$ (it could be $L^2[0,1]$ or a reproducing kernel Hilbert subspace thereof) that manifest different kinds of dispersion around their mean. The functions could arise from bio-physical applications as in \citet{panaretos2010second},
\citet{kraus2012dispersion}, and \citet{tavakoli:2014}, where functional
curves represent DNA mini-circles; from linguistic applications, where the
functional curves represent the spectrograms of spoken phonemes by different
speakers (see e.g. \citet{pigoli2018statistical}), and one may wish to classify several groups of sounds; from the analysis of wheel-running
activity curves in mice, where one may wish to distinguish several levels of activity evolved across age and generation selections (\citet{cabassi2017permutation}); or again from neuroscience experiments recording sequential trajectories where each trajectory consists of oscillations and fluctuations around zero (\citet{jiaoshuaofrostigombao2022}). What these examples have in common is that it is not the mean structure of the curves that captures the most interesting differences; rather, their (dis)similarities are nested into their second-order structure, embodied by the functional covariance operators. 

First-order functional clustering, that is, the clustering of observations based on their mean structure, is a
topic widely examined in the literature (see, e.g. the surveys by
\citet{jacques2014functional}, \citet[Section 2.7]{aneiros2022functional}, and more pertinent to the scope of this paper, \citet{chamroukhi2019model} for model-based clustering). Less literature
exists on the classification of covariance operators. 
Clustering and classifying functional covariances
require a way to compare and quantify differences between them, which
makes the task more challenging. 

Work in this direction can predominantly be found in the multivariate finite dimensional setting, where several methods for clustering covariance matrices exist, especially in connection with statistical shape analysis and diffusion tensor
imaging (see, among others,
\citet{lee2015geodesic} and \citet{srivastava2016functional}). Only recently, methods for
classifying infinite-dimensional covariance operators gained some traction. 
\citet{ieva2016cluster} consider differences between global covariance
structures limited to the case of two classes of equal size. They focus
specifically on a set of observations from two populations who exhibit
only second-order variation (same mean, different covariances).
\citeauthor{jiaoshuhau2020}'s \citep{jiaoshuhau2020}
Hilbert--Schmidt based classification is driven by a biomedical
(neurological) application. \citet{peng2018cluster} use a covariance
dissimilarity measure to classify self-similar stochastic processes when
the number of clusters is known, and they later expand it in
\citet{peng2019covariance}, developing an asymptotically consistent
algorithm for classifying of wide-sense stationary ergodic
processes.

What the works above have in common is that they measure
similarities by embedding covariances in a much larger, linear
Hilbert--Schmidt space, and they employ the relative Hilbert--Schmidt
(Frobenius) distance. However, covariance operators can essentially be seen as squares of Hilbert--Schmidt
operators (\citet{pigoli2014distances,masarotto2018procrustes}), and as such, the space they inhabit is intrinsically non-linear. Although
such non-linearity increases the difficulty of the task, properly accounting for it does yield more
statistical power. \citet{pigoli2014distances} and 
\citet{masarotto2018procrustes} argue that the Wasserstein-Procrustes distance
is the most natural metric that allows to perform statistical analysis on covariance operators, while at the same time respecting their non-linear nature.  In this paper, we
employ a framework based on the Wasserstein distance of Optimal Transport in order to cluster data with respect to
differences in their covariance structure.
The Wasserstein framework was recently employed for data
clustering in \citet{verdinelli2019hybrid}. The authors perform hard
distribution clustering with a modified version of the Wasserstein
distance and provide successful numerical simulation in dimensions 1 and
2, where they also estimated the number of groups.
\citet{kashlak2019inference} also developed a methodology that accounts for non-linearity and
employ an expectation-maximization algorithm to separate curves or
operators into $k$ sets via a concentration of measure procedure. They
propose an algorithm that updates a vector of probabilities for each
operator to belong to a certain cluster until a local optimum is
reached. When given the correct number of clusters, their method outperforms standard $k$-means clustering, especially in the
highly difficult case of classifying rank 1 operators. 

A potential common limitation of the aforementioned methods 
is that, with the partial exception of \citet{kashlak2019inference},  
specialize in hard clustering; that is, each observation is classified in only one group. 
In contrast, we propose a fuzzy clustering algorithm (referred to as soft clustering in the following)
directly based on the Wasserstein-Procrustes distance, in which the in-between cluster variability is
penalised by a term proportional to the entropy of the partition matrix. The algorithm characterizes
each group using a centroid covariance operator, and simultaneously estimates the degree to which the observed 
covariances belong to each cluster.
Soft classification algorithms resemble model-based
clustering, which are techniques that rely on the underlying
assumption that the data come from a mixture of distributions (\citet{fraley2002clustering}).
However, the estimated membership grades do not have a precise probabilistic interpretation, because soft-clustering is not based on any stringent probabilistic assumption.
 Unlike $k$-means and other hard clustering methods,
soft clustering is particularly
meaningful in cases in which the cluster separation is not neat, and a soft
classification,  which allows for overlapping clusters,  might be appropriate. 
Furthermore, it allows us to easily identify the groups that are most confused with each other.
We also attempt to ameliorate some of the intrinsic weaknesses of related works
by (a) proposing an easy adaptation of the methodology that supports fast clustering when the number of covariance operators is very large and (b) offering insights into how to determine the most suitable number of
clusters. Almost all the works previously described require in fact the number
of clusters to be known in advance. We introduce a novel cluster quality index, coined Trimmed Average Silhouette Width (TASW), which provides an attractive alternative to the well-known Average Silhouette Width (\citet{rousseeuw_silhouettes_1987}) in the case of overlapping clusters and shows extremely good performances in the applications presented. Moreover, the TASW may also be used to test the null hypothesis of whether any cluster is present. Finally, we also aim to fill what, to the
best of our knowledge, is a gap  in the field: the lack of ready-made,
easy-to-use clustering code. 

The structure of the paper is as follows.
After introducing some basic notions and notations in Section~\ref{sec:background},
we describe the proposed soft clustering approach in Section~\ref{sec:soft_clustering}. 
In particular, in Section~\ref{sec:soft_clustering}, we discuss the methodology, the computational details (also for large datasets), and a cluster quality index that we have found to be useful for selecting the number of groups.
In Section~\ref{sec:numerical}, we illustrate the usefulness of the procedure 
on simulated data, as well as two real datasets. 
Conclusions and some ideas for future research are presented in Section \ref{sec:conclusion}.
An \texttt{R} implementation of our algorithm 
is given in the supplementary materials together with an \text{R} script that allows to reproduce part of one of the examples in Section~\ref{sec:numerical}.

\section{Background}
\label{sec:background}
\subsection{Basic notions and notations}
\label{subsec:notations}

Let $\mathcal{H}$ be a real separable Hilbert space, often taken as $L^2[0,1]$, with the inner product
$\langle\cdot,\cdot\rangle:\mathcal{H}\times\mathcal{H}\rightarrow\mathbb{R}$, and
induced norm $\|\cdot\|:\mathcal{H}\to[0,\infty)$.
Furthermore, let $\{X_{1,j}\}_{j=1}^{n_1}, \dots ,\{X_{N,j}\}_{j=1}^{n_N}$ be $N$
independent samples of i.i.d. random elements in $\mathcal{H}$,
possessing, for $i=1,\ldots,N$,  well-defined mean functions $\mu_i = E\bigl\{X_{i,j}\bigr\}$ and
covariance operators $\Sigma_i = E\bigl\{(X_{i,j}-\mu_i)\otimes(X_{i,j}-\mu_i)\bigr\}$, 
where $\otimes$ stands for the outer product on $\mathcal{H}$.
With this notation, we assume that the data might arise from $N$
functional populations modelled by a prototypical random function $X_i$,
and that we are able to observe $n_i>1$ realisations from each population. 
Populations might differ in both mean and covariance structure. However, as previously mentioned, we assume that the scientific interest concerns the differences in the covariance operators.

We  say that a (possibly unbounded) operator $A$ is
\emph{self-adjoint} if $\langle Au,v\rangle = \langle u,Av\rangle$ for
all $u$ and $v$ in the domain of definition of $A$. If $A$ also happens to be
bounded, then this is equivalent to the condition that $A=A^*$,
where $A^*$ denotes the adjoint operator.
A \emph{non-negative} operator is a self-adjoint
operator $A$ such that $\langle Au,u\rangle\ge0$ for all $u$ in the
domain of $A$.  In addition, if $A$ is compact, then there exists a
unique non-negative self-adjoint 
operator whose square equals $A$, which is
denoted by either $A^{1/2}$ or $\sqrt{A}$.  The inverse square root
$(A^{1/2})^{-1}$ is denoted by $A^{-1/2}$.  The inverse may not be
defined or defined only on a (dense) subspace of $\X$. For any bounded
operator $A$, $A^*A$ is non-negative.
The identity operator on $\mathcal{H}$ is denoted by $\mathscr{I}$.


In this setting, covariances are linear operators from $\mathcal{H}$
into $\mathcal{H}$, which are \emph{self-adjoint}, \emph{non-negative},
and \emph{trace-class}, meaning that their nuclear norm $\hs A
\hs_1=\tr\sqrt{A^*A}$, with $\tr$ denoting the trace operator,  
is bounded away from infinity. 
Thus, a covariance operator $\Sigma$ on $\mathcal{H}$ can be considered
the ``square" of a Hilbert--Schmidt operator: 
If $\hs A\hs_2 = \sqrt{\tr A^*A} < \infty$,
then $A$ is certainly bounded, and $\Sigma=A^*A$ defines a valid covariance
operator. Viewing covariances as squares highlights their non-linear
nature, and any method for measuring (dis)similarities should account for
this non-linearity. 

\subsection{On the choice of a suitable distance}
\label{subsec:metric}

A classical strategy for dealing with non-linearity in infinite dimension has been
embedding covariances in a much larger, linear Hilbert-Schmidt space,
and comparing them directly by means of the Hilbert-Schmidt distance $\hs
\Sigma_1-\Sigma_2\hs_2=\sqrt{\tr(\Sigma_1-\Sigma_2)^*(\Sigma_1-\Sigma_2)}$
(\citet{panaretos2010second,fremdt2013testing,boente2018testing}).
However, \citet{pigoli2014distances} showed that there exists a much more
natural metric that respects the non-linear nature of covariances
while being well-defined in infinite dimensions, that is,  the
so-called \emph{Procrustes metric} 
\[
\Pi(\Sigma_1,\Sigma_2)=\inf_{U^*U=\mathscr{I}}\hs
\Sigma^{1/2}_1-\Sigma^{1/2}_2U \hs_2 .
\] 
\citet{masarotto2018procrustes} further developed several key properties of this metric and its
geometry. Their development leverages  the observation that two
covariance operators $\Sigma_1$ and $\Sigma_2$ can be
bijectively identified with two centered Gaussian measures
$N(0,\Sigma_1)$ and $N(0,\Sigma_2)$ on the Hilbert space
$\mathcal{H}$. The Procrustes metric can then be interpreted via the
optimal transportation of Gaussian processes, and identified with the
$L^2$-Wasserstein distance between the two Gaussian measures.

The Wasserstein distance between two Borel probability measures $\mu$ and $\nu$,
both on $\mathcal H$,   is defined as 
\[ W^2(\mu,\nu) =\inf_{\pi\in
\Gamma(\mu,\nu)} \ownint{\mathcal H\times\mathcal
H}{}{\|x-y\|^2}{\pi(x,y)},
\] 
where the infimum is taken over the set
$\Gamma(\mu,\nu)$ of all the couplings of $\mu$ and $\nu$. 
Alternatively, 
\[
W^2(\mu,\nu) = \inf_{(Z_1, Z_2)}\mathbb E\|Z_1 - Z_2\|^2,
\] 
where the infimum is over all random vectors $(Z_1,Z_2)$ in $\mathcal
H\times \mathcal H$ such that $Z_1 \sim \mu$ and
$Z_2 \sim \nu$, marginally.
The distance is
finite when $\mu$ and $\nu$ have a finite second moment. The 
optimisation problem above is known as the Monge--Kantorovich problem of
optimal transportation. 
If $\mu$ and $\nu$ are Gaussian measures $N(0,\Sigma_1)$ and
$N(0,\Sigma_2)$, then the Wasserstein distance can be expressed in
closed form as
\begin{align*}
W^2(\mu, \nu) 
&= \Pi^2(\Sigma_1,\Sigma_2) = 
\inf_{\begin{smallmatrix} Z_1\sim N(0,\Sigma_1)\\Z_2 \sim N(0, \Sigma_2)\end{smallmatrix}} \mathbb{E}\|Z_1-Z_2\|^2 \\
     &=\tr\Sigma_1+\tr\Sigma_2-2\tr\sqrt{\Sigma_1^{1/2}\Sigma_2\Sigma_1^{1/2}}.\\
\end{align*}


\subsection{Weighted Fr\'echet means/Wasserstein barycenters}
\label{subsec:frechet}

If  $\Sigma_1,\dots,\Sigma_N$ are covariance operators, and $\pi_1,\ldots,\pi_N$ are non-negative weights, 
their (weighted) Fr\'echet mean (\citet{frechet1948elements}) with respect
to the Procrustes metric  (or equivalently, their Wasserstein barycenter of
the corresponding centred Gaussian measures) is defined as the minimiser of the 
Fr\'echet functional
\[ 
F(\Sigma) =
\sum_{i=1}^N  \pi_i \Pi^2(\Sigma_i,\Sigma) =
\sum_{i=1}^N \pi_i W^2\bigl(N(0,\Sigma_i),N(0,\Sigma)\bigr). 
\]  
Unlike the finite dimension case, the existence and
uniqueness of Fr\'echet means in general metric space is not guaranteed
(see, e.g., \citet{karcher1977riemannian}).  In Wasserstein spaces, 
however, these can be established under rather mild assumptions thanks
to the notion of optimal multicoupling
(\citet{masarotto2018procrustes}). In particular, any collection
$\Sigma_1,\dots,\Sigma_N$ of covariance operators admits a Fr\'echet
mean $\overline\Sigma$ with respect to the Procrustes distance $\Pi$
which is stable under finite dimensional projections. For Gaussian
measures, it is uniquely defined, and is a Gaussian measure itself (see
\citet{agueh2011barycenters}). 

There is no closed-form formula that returns the Fr\'echet
mean $\overline\Sigma$, but as
described, for example,  in \citet[Section~8]{masarotto2018procrustes},  it can be iteratively approximated
using the following algorithm.

\begin{description}
\item[Initialization.] Set $j=0$ and $\overline\Sigma^0$ equal to an initial guess of the Fr\'echet mean
(e.g., $\overline\Sigma^0 =\sum_{i=1}^N \pi_i \Sigma_i / \sum_{i=1}^N \pi_i$).
\item[Computation.]
Until a suitable convergence criterion is met (or a maximum number of iterations is reached), repeat
\begin{itemize}
    \item[--] Compute the next iterate as \(\overline\Sigma^{j+1}=\overline T_j\overline\Sigma^j \overline T_j\)
    where
    \[
    \overline T_j = \left(\sum_{i=1}^N \pi_i T_{i,j}\right)/\left(\sum_{i=1}^N \pi_i\right)
    \]
    with
\[
T_{i,j}
=\left(\overline\Sigma^j\right)^{-1/2}
\left[\left(\overline\Sigma^j\right)^{1/2}\Sigma_i\left(\overline\Sigma^j\right)^{1/2}\right]^{1/2}
\left(\overline\Sigma^j\right)^{-1/2} \;\;(i=1,\ldots,N).
\]
\item[--] Set $j=j+1$
\end{itemize}
\item[Output.] Use the final iterate $\overline\Sigma^j$ as an estimate of $\overline\Sigma$.
\end{description}

Note that $T_{i,j}$ is a self-adjoint operator that \emph{transports} $\overline\Sigma^j$ to $\Sigma_i$, in the sense that $\Sigma_i=T_{i,j}\overline\Sigma^J T_{i,j}$.  In terms of the manifold-like geometry of covariances under the Wasserstein-Procrustes metric, the algorithm starts with an initial guess of the Fr\'echet mean. It then lifts all the covariances to the tangent space at that initial guess, averages linearly on the tangent space,  and retracts this average onto the manifold. This retraction is the guess in the following step. 
The previous algorithm can also be viewed as an implementation of the steepest descent concept in the space of covariances endowed with the Wasserstein-Procrustes metric (\citet{zemel2019frechet,masarotto2018procrustes}). 

\subsection{Finite-dimensional behaviour}

In applications such as those presented in Section \ref{sec:numerical}, we inevitably work with  finite-dimensional representations of the covariance operators. However, as discussed in  \citet{pigoli2014distances} and  
\citet{masarotto2022procrustes}, the Wasserstein-Procrustes distance between two finite-dimensional representations provides a good approximation of the corresponding distance between the infinite-dimensional operator.
In addition, \citet{masarotto2022procrustes} established the numerical stability of the Wasserstein barycenters.

\section{Soft clustering of covariance operators}\label{sec:soft_clustering}

\subsection{Generalities}
\label{subsec:generalities}

As previously mentioned, assume that 
\begin{itemize}
    \item  we have observed $N$ independent samples, each of size $n$, of functional data 
    $\{X_{1,j}\}_{j=1}^{n_1}, \dots ,\{X_{N,j}\}_{j=1}^{n_N}$;
    \item we wish to cluster the corresponding covariances $\Sigma_i$, $i=1,\ldots, N$,
    in $K$ groups.
\end{itemize}
In practice, we have $N$ estimated operators
$\widehat{\Sigma}_1,\ldots,\widehat{\Sigma}_N$, for example, the sample covariances defined by
\[
\widehat\Sigma_i = \dfrac{1}{n_i-1} 
\sum_{j=1}^{n_i} \bigl(X_{i,j}-\widehat\mu_i\bigr)\otimes\bigl(X_{i,j}-\widehat\mu_i\bigr)
\text{ where } \widehat\mu_i = \dfrac{1}{n_i} \sum_{j=1}^n X_{i,j},
\]
and we aim to determine
\begin{itemize}
\item[(a)] 
$K$ prototype covariance operators $\overline{\Sigma}_1,\ldots,\overline{\Sigma}_K$
representative of the $K$ groups; and 
\item[(b)]
a $N \times K$ (\emph{soft}) partition matrix 
\[
P=[\pi_{i,j}] \text{ such that } \pi_{i,j} \ge 0 \text{ and }
\sum_{j=1}^K \pi_{i,j} = 1, 
\]
where each element $\pi_{i,j}$ describes the confidence
with which the covariance $\widehat{\Sigma}_i$ can be assigned to the $j$th
group. 
\end{itemize}
A natural scenario would see the $j$th cluster barycenter $\overline\Sigma_j$
be the Fr\'echet means of the sample covariances with weights proportional to the membership grades $\pi_{1,j},\ldots,\pi_{N,j}$ and to the sample sizes $n_1, \ldots, n_N$;
simultaneously the grades $\pi_{i,j}$ should be determined so that they are close to one when $\widehat\Sigma_i$ is \emph{near}
$\overline\Sigma_j$, and close to zero when $\widehat\Sigma_i$ looks like an \emph{outlier} for the group identified by $\overline\Sigma_j$. However, we also recognize that not all the operators have a clear-cut attribution to one of the groups, and with the aim of identifying them and reducing their influence in the barycenters computation, we let $\pi_{i,j}$ assume any value between zero and one. 

In particular, the proposed soft clustering method computes
$\overline{\Sigma}_1,\ldots,\overline{\Sigma}_K$ and $P$ minimizing
\begin{align}
  &\sum_{i=1}^N\sum_{j=1}^K \pi_{i,j}\bigl(n_i-1\bigr)\Pi^2\bigl(\widehat{\Sigma}_i,\overline{\Sigma}_j\bigr)
\label{eqn:obj} \\
\intertext{subject to the constraints}
  &\pi_{i,j} \ge 0\;\; (i=1,\ldots,N \text{ and } j=1,\ldots,K) \tag{C1}\label{eqn:pi+},\\
  &\sum_{j=1}^K \pi_{i,j} = 1\;\;(i=1,\ldots,N)\tag{C2}\label{eqn:sum1},\\
  -\dfrac{1}{N}&\sum_{i=1}^N\sum_{j=1}^K \pi_{i,j}\log\pi_{i,j}=E, \tag{C3}\label{eqn:average-entropy}
\end{align}
where $E$ is a user-defined value ($0 \le E \le \log K$). 

The objective function \eqref{eqn:obj}, given by the weighted sum of the Wasserstein-Procrustes distances, measures the heterogeneity 
within the classes, while the constraint \eqref{eqn:average-entropy} prescribes the desired average level of entropy of the
resulting soft classification. In particular, observe that
\begin{itemize}
\item[(i)]
when $E=0$ (null average entropy), the constraint can be satisfied only by standard (or \emph{hard}) partition matrices such that,
for every $i$,
\begin{equation}
\pi_{i,j} = 
\begin{cases}
  0, & j \ne r_i\\
  1, & j=r_i
\end{cases}
\label{eqn:noentropy}
\end{equation}
for some $r_i \in \{1,\ldots,k\}$.
\item[(ii)]
when $E=\log K$ (maximum average entropy), the constraint implies the non-informative partition matrix 
\begin{equation}
\pi_{i,j} = \dfrac{1}{K}\;\; (i=1,\ldots,N;\; j=1,\ldots,K).
\label{eqn:uniform}
\end{equation}
\item[(iii)]
when $0 < E < \log K$, the rows of the resulting partition matrix will neither degenerate as in  \eqref{eqn:noentropy} nor become uniform as in \eqref{eqn:uniform}.
\end{itemize}

In the following subsections, we discuss 
the computation of the soft cluster solution 
(Subsection \ref{subsec:comp_impl});
the choice of the average entropy $E$ (Subsection \ref{subsec:E});
an index, the trimmed average silhouette width, that can be used
for selecting suitable values for $K$ (Subsection \ref{subsec:K}); and
a simple adaptation of our approach to the ``large $N$'' scenario (Subsection \ref{subsec:large}).

\subsection{Computation of the soft clustering solution}
\label{subsec:comp_impl}

The following proposition, whose proof is sketched in the appendix, naturally leads to the algorithm we implemented 
for computing the cluster barycenters $\overline\Sigma_1,\ldots,\overline\Sigma_K$ and the partition matrix $P$.
\begin{proposition}
In the previously described setting
\begin{itemize}
\item[1.] given the partition matrix $P$, 
    the desired covariance matrices/cluster barycenters $\overline{\Sigma}_j$, $j=1,\ldots,K$, are
    the Fr\'echet means of  $\widehat\Sigma_1, \ldots, \widehat\Sigma_N$ with weights $(n_1-1)\pi_{1,j}, \ldots, 
    (n_N-1)\pi_{N,j}$, 
    i.e., 
    \begin{equation}
      \overline{\Sigma}_j 
      = \arg \min_{\Omega} \sum_{i=1}^N \pi_{i,j} \bigl(n_i-1\bigr)\Pi^2\bigl(\widehat{\Sigma}_i,\Omega\bigr);
    \label{eqn:clbary}
    \end{equation}
\item[2.] given $\overline{\Sigma}_j$, $j=1,\ldots,K$, the partition matrix that minimizes
    \eqref{eqn:obj} under the constraints \eqref{eqn:pi+}--\eqref{eqn:average-entropy} is
    \begin{equation}
    P=\bigl[\pi_{i,j}(\widehat\eta)\bigr]
    \label{eqn:poptimal}
    \end{equation}
    where
    \begin{equation}
    \pi_{i,j}(\eta)=
    \dfrac{e^{-(n_i-1)\Pi^2(\widehat{\Sigma}_i,\overline{\Sigma}_j)/\eta}}
    {\sum_{s=1}^Ke^{-(n_i-1)\Pi^2(\widehat{\Sigma}_i,\overline{\Sigma}_s)/\eta}},
    \label{eqn:pformula}
    \end{equation}
    and $\widehat\eta$ denotes the unique positive root of the equation
    \begin{equation}
        -\sum_{i=1}^N \sum_{j=1}^K \pi_{i,j}(\eta) \log \pi_{i,j}(\eta) = NE.
        \label{eqn:etaconstraint}
    \end{equation}
    In addition, the left side of \eqref{eqn:etaconstraint} is differentiable and monotone increasing when $\eta>0$. Thus, the computation of $\widehat\eta$ is a stable, and essentially trivial, numerical problem.
\end{itemize}
\end{proposition}

The result motivates the use of the block coordinate descent algorithm (e.g., \citet{xu2013block})
described in Subsection \ref{subsub:blk}. The algorithm, which resembles the classical EM approach for fitting mixture models 
(and model-based clustering), finds only a local minimum of the
objective function \eqref{eqn:obj}. Thus, it is important to choose a suitable starting point 
so that this local minimum corresponds to a ``good'' solution. With this aim, and inspired by
the initialisation phase of the \texttt{kmeans++} (\citet{vassilvitskii2006k})
and \texttt{PAM} (\citet{kaufman2009finding}) algorithms, we suggest the initialization approach
described in Subsection \ref{subsub:ini}.

\subsubsection{Initialisation}
\label{subsub:ini}

During the initialization phase, we try to minimize the sum of the within-group distances \eqref{eqn:obj}
adding to the constraints \eqref{eqn:pi+}--\eqref{eqn:average-entropy} 
the additional restriction that  the group prototypes are equal to some of the observed sample covariances.
Therefore, we try to determine the indices $(i_1,\ldots,i_K) \in \{1,\ldots,N\}^K$ that minimize
\[
G(i_1,\ldots,i_K)=\sum_{i=1}^N\sum_{j=1}^K \pi_{i,j}(\widehat\eta) 
\bigl(n_i-1\bigr)\Pi^2\bigl(\widehat\Sigma_i, \widehat\Sigma_{i_j}\bigr)
\]
with $\widehat\eta$ and $\pi_{i,j}(\widehat\eta)$ computed as in the previous proposition assuming that
\( \overline{\Sigma}_1 = \widehat{\Sigma}_{i_1},\ldots,\overline{\Sigma}_K = \widehat{\Sigma}_{i_K}\).
Restricting the search to the observed sample covariances avoids  
the computing of the Fr\'echet means,  as this is the most time-consuming part of the 
algorithm described in the next Subsection. As the exact determination of the optimal subset $(i_1,\ldots,i_K)$
is not computationally feasible (at least when $N$ is large), 
we suggest the use of the following stochastic search approach:
\begin{enumerate}
  \item[] Repeat \texttt{nstart} times the following steps  and
    keep the best subset $(i_1,\ldots,i_K)$ generated.
  \begin{enumerate}  
    \item[--] Choose $i_1$ uniformly at random in $\{1,\ldots,N\}$. 
    \item[--] For $j=2,\ldots,K$, sequentially choose
      $i_j$ from $\{1,\ldots,N\}$ with probability proportional
      to 
      $\min\bigl\{\Pi^2(\widehat{\Sigma}_i,\widehat{\Sigma}_{i_1}),\ldots,\Pi^2(\widehat{\Sigma}_i,\widehat{\Sigma}_{i_{j-1}})\bigr\}$.
    \item[--] Repeat \texttt{nrefine} times the following step. 
    \begin{enumerate}
      \item[] For each $j=1,\ldots,K$, sample without replacement from $\{1,\ldots,N\}$ 
        \texttt{ntry} possible substitutions
        of $i_j$  with probability proportional to
        $ \min_{s \ne j} \Pi^2(\widehat{\Sigma}_i,\widehat{\Sigma}_{i_s}).$
        Keep the best found value for $i_j$.
      \end{enumerate}
    \end{enumerate}
  \end{enumerate}
The algorithm described above is based on the idea that  if $i_a$ and $i_b$ belong to $\{i_1,\ldots,i_K\}$ then the distance $\Pi^2(\widehat{\Sigma}_{i_a},\widehat{\Sigma}_{i_b})$ is expected to be large.   
Repeated simulations showed that \emph{if} $K$ groups really exist,  this initialisation tends to select covariances
$\widehat{\Sigma}_{i_1},\dots,\widehat{\Sigma}_{i_K}$ belonging to
different groups, at least when \texttt{nstart} and \texttt{nrefine} are greater than,  or equal,  to $5$, and \texttt{ntry} is about $N/K$. 

\subsubsection{Block coordinate descent algorithm}
\label{subsub:blk}

The local search follows directly from the given proposition. The algorithm starts from the $K$ prototype covariance matrices
$\widehat\Sigma_{i_1},\ldots,\widehat\Sigma_{i_K}$ selected during the initialization phase and then
seeks the solution to the soft cluster problem through iteration of the following two steps until convergence:
\begin{enumerate}
  \item
    Given the current estimates of the cluster barycenters, compute the optimal partition matrix from 
    \eqref{eqn:poptimal}--\eqref{eqn:etaconstraint}.
  \item 
    Given the current partition matrix, update the  $K$ prototype covariance matrices 
    using the gradient descent algorithm for the Fr\'echet mean (see Subsection~\ref{subsec:frechet}).
  \end{enumerate}
The algorithm stops when the difference between the sum of the within-class distances \eqref{eqn:obj} in two consecutive iterations  is sufficiently small.

\subsection{On the choice of the average entropy $E$}
\label{subsec:E}

In general, $E$ should be chosen based on the the expected \emph{average degree of confusion} between the clusters. In practice, we obtained good results by setting 
\begin{equation} 
E =  -\bigl(1-\alpha\bigl)\bigl[ \beta \log \beta + (1-\beta) \log (1-\beta) \bigr]+\alpha \log 2 
\label{eqn:suggestedE}
\end{equation}
for positive small values of $\alpha$ and $\beta$ (e.g., we used 
$\alpha=0.25$ and $\beta=0.05$ in all the applications presented in Section \ref{sec:numerical}). 
The rationale behind \eqref{eqn:suggestedE} is that, in general, we expect
\begin{itemize}
\item[(i)]
most of the sample covariances (say $100(1-\alpha)\%$ of them) to be classified essentially in one group 
(denote it with $r_{i}$), with some uncertainty about another group ($s_{i}$) ; a prototypical partition matrix row for these cases is
$$
\pi_{i,j} = \begin{cases}
  1-\beta & \text{if } j=r_{i} \\
  \beta   & \text{if } j=s_{i} \\
  0       & \text{otherwise}
\end{cases}
.
$$  
\item[(ii)]
the remaining small number of sample covariances ($100\alpha\%$) to be \emph{confused} between two possible overlapping clusters
($r_i$ and $s_i$); for these cases, a prototypical row of $P$ is
$$
\pi_{i,j} = \begin{cases}
  \dfrac{1}{2} & \text{if } j \in \{r_{i}, s_{i}\} \\
  0       & \text{otherwise}
\end{cases}
.
$$
\end{itemize}
Of course, we do not expect any rows of $P$ to be exactly equal to the previous prototypical rows, but at least in our experience, 
the use of \eqref{eqn:suggestedE} makes it easy to specify a suitable value for the average entropy.

\subsection{Trimmed average silhouette width}
\label{subsec:K}

In the standard multivariate framework, many cluster quality indexes have been proposed, in particular,  with the aim of determining a suitable value for the number of clusters. One popular index is the average silhouette width proposed by \citet{rousseeuw_silhouettes_1987},  which achieved very good results in \citeauthor{arbelaitz_extensive_2013}'s \citep{arbelaitz_extensive_2013}
extensive study, in which $30$ different indexes were compared.  
This criterion has been considered not only for choosing an optimal value of $K$ but also for obtaining good partitions for a fixed value of $K$ (see \citet{van_der_laan_new_2003, batool_clustering_2021}). 

As the silhouette width is a widely used intuitive measure of cluster validity, we explored a simple adaptation of the idea to our framework with the aim of providing a tool that can 
suggest ``good'' values for the number of cluster $K$, and if desired, is usable to test for the presence of any cluster, that is, the null hypothesis that $K=1$.  

Assume that we have computed the soft cluster solutions for $K=2,\ldots,K_{max}$, all using the same average entropy $E$, and denote the membership grades with $\pi_{i,j}^K$ and the cluster barycenters with $\overline\Sigma_j^K$, $i=1, \ldots, N$ and $j=1,\ldots,K$. 
We define the silhouette width of the $i$th sample covariance $\widehat\Sigma_i$ with respect to the solutions based on $K$ clusters as
\begin{equation}
SW_i^K = 1-\dfrac
{\Pi\bigl(\widehat\Sigma_i, \overline\Sigma_{i_a^K}^K\bigr)}
{\Pi\bigl(\widehat\Sigma_i, \overline\Sigma_{i_b^K}^K\bigr)},
\label{eqn:silhouette}
\end{equation}
where $\overline\Sigma_{i_a^K}^K$ and $\overline\Sigma_{i_b^K}^K$ are the nearest and second nearest 
barycenters to $\widehat\Sigma_i$, that is, 
\[
i_a^K = \arg \min_j \Pi\bigl(\widehat\Sigma_i, \overline\Sigma_j^K\bigr)
\text{ and }
i_b^K = \arg \min_{j \ne i_a^K} \Pi\bigl(\widehat\Sigma_i, \overline\Sigma_j^K\bigr).
\]
Intuitively, $SW_i^K$ measures how well the $i$th sample covariance can be classified in the
cluster identified by the nearest barycenter with respect to the second best classification.
Observe that our definition of $SW_i^K$ is analogous to the fast silhouette definition considered by
\citet{van_der_laan_new_2003}. With respect to the original definition given by \citet{rousseeuw_silhouettes_1987}, \eqref{eqn:silhouette} offers the advantage of requiring only quantities already computed during the computation of the soft cluster solution. 

The average silhouette width considered in the literature is the mean of the individual silhouettes $SW_i^K$, that is, $ASW^K=\sum_{i=1}^N SW_i^K/N$. However, this index is not completely appropriate in the framework considered in this paper, because we \emph{a priori} accept the idea that clusters might overlap, and therefore, that some of the individual silhouette width could be small even for a ``good'' partition. 
However, in a ``good'' soft classification, 
\begin{equation}
\text{C}_i^K = \pi_{i,i_a} = \max_{j=1, \ldots, K} \pi_{i,j}^K
\label{eqn:credibility}
\end{equation}
should be able to measure the credibility of the classification of the $i$th covariance to its nearest barycenter, and, in particular, we wish the silhouette widths $SW_i^K$ to be large when $C_i^K$ is large. This heuristic reasoning leads us to define the
\emph{trimmed average silhouette width}
\[
TASW_K = 
\dfrac
{\sum_{i \in \text{GOOD}}(n_i-1)SW_i^K}
{\sum_{i \in \text{GOOD}} (n_i-1)}
\]
where 
$$\text{GOOD}=\bigl\{i: C_i^K \ge \sum_{i=1}^N C_i^K /N\bigr\}.$$

The index $TASW_K$ measures how well the ``core'' part of each group is homogeneous and well separated from the other groups. Thus, in general, large values of $TASW_K$ should point to reasonably good classifications. In particular, 
\[
\widehat{K} = \arg \max_{K=2,\ldots,K_{max}} TASW_K 
\]
can be used to estimate the optimal number of clusters. More generally, recognising the uncertainty of
every single choice of $K$, the set
\[
\bigl\{K: (TASW_{max}-TASW_{K})/TASW_{max} \le \delta \bigr\}
\]
should be explored. Here, $\delta$ denotes a small positive number (say $0.05$),  and
\[
TASW_{max} = \max_{K=2,\ldots,K_{max}} TASW_K.
\]

In addition, observe that as we  illustrate in the next section,  
$TASW_{max}$ can be used to test the null hypothesis of a single cluster ($K=1$) against the alternative hypothesis of $K>1$ clusters. An approximate reference null distribution can be obtained using a permutation approach, that is, by repeatedly randomly rearranging the $\sum_i n_i$ centered observations $X_{i,j}-\widehat\mu_i$ in $N$ samples of size $n_1, \ldots, n_N$ and recomputing the $K_{max}$ classifications and the corresponding $TASW_{max}$ statistic. 

\subsection{Clustering a large number of covariance operators}
\label{subsec:large}

The computation complexity of the algorithm in  Subsection \ref{subsec:comp_impl} increases only linearly in the number of covariance operators $N$. However, computing a single Wasserstein-Procrustes distance requires $\mathcal{O}\bigl(M^3\bigr)$ operations, 
where $M \times M$ is the size of the matrices used as finite approximations of the infinite-dimensional covariance operators.
Thus, the computation of the cluster solution can be slow when $N$ and $M$ are large. A possibility for reducing the computational burden consists of using finite-dimensional approximations with a low resolution, that is, a smaller value for $M$. However, in this way, we can lose some important details of the sample covariances. For this reason, when $N$ is large,  we suggest 
proceeding in the following way:

\begin{enumerate}
    \item 
    Estimate the cluster barycenters $\overline\Sigma_1, \ldots, \overline\Sigma_K$ by applying the algorithm in Subsection \ref{subsec:comp_impl} to a subset of $N_{reduced}$ randomly chosen sample covariances.
    \item
    Compute the full partition matrix following \eqref{eqn:poptimal}--\eqref{eqn:etaconstraint} from the previously determined barycenters.
\end{enumerate}

As we illustrate in the next section, using this simple approach,  
we were able to cluster thousands of covariance operators.
Naturally, if desired and the computational resources permit, the two steps can be repeated a number of times, keeping the best solutions, to reduce the risk that not all groups are well represented in the subset used during the first step. 

\section{Numerical experiments}\label{sec:numerical}

\subsection{Synthetic data}
In this subsection, we summarize the results obtained by applying our clustering algorithm 
to simulated data.  In the scenario, there are four clusters, and in particular, 
we considered functional data simulated,  for $u \in [0,1]$, as
\[
X(u) = \begin{cases}
\sum_{r=0}^{32} \lambda^r \xi_{r} f_r(u) + \zeta f_1(u)     & \text{first cluster,} \\
\sum_{r=0}^{32} \lambda^r \xi_{r} f_r(u) + \zeta f_2(u)     & \text{second cluster,}\\
\sum_{r=0}^{32} \lambda^r \xi_{r} f_r(u) + \zeta f_3(u)     & \text{third cluster,}\\
\sum_{r=0}^{32} \lambda^r \xi_{r} f_r(u) + \zeta f_4(u)     & \text{fourth cluster,} 
\end{cases}
\]
where 
\begin{itemize}
    \item $\lambda=2/\sqrt{5}$;
    \item $\xi_{r}$ and $\zeta$ are independent standard normal random variables; 
    \item $f_r(u)$ are the elements of the orthonormal Fourier basis on the unit interval, that is, 
    \[
    f_r(u) = \begin{cases}
      1 & \text{if } i=0, \\
     \sqrt{2}sin\bigl((i+1) \pi  u\bigr) & \text{ if } i=1, 3, \ldots \\
     \sqrt{2}cos\bigl(i \pi u\bigr) & \text{ if } i=2, 4, \ldots
    \end{cases}
    .
    \]
\end{itemize}

Therefore, 
\[
X \sim \begin{cases}
N(0, \Sigma+\Delta_1)     & \text{first cluster}, \\
N(0, \Sigma+\Delta_2)     & \text{second cluster}, \\
N(0, \Sigma+\Delta_3)     & \text{third cluster}, \\
N(0, \Sigma+\Delta_4)     & \text{fourth cluster}, 
\end{cases}
\]
where 
\[
\Sigma = \sum_{r=0}^{32} \lambda^{2r} f_r \otimes f_r \text{ and } \Delta_j = f_{j} \otimes f_{j}\;(j=1,\ldots, 4),
\]
and $\otimes$ denotes the outer product in $L^2[0,1]$.

The cluster algorithm was applied to
\begin{itemize}
\item[(i)]
$100$ datasets consisting of $N=100$ sample covariances, $25$ for each cluster; for these datasets, 
we used the ``full'' algorithm described in Subsection  \ref{subsec:comp_impl};
\item[(ii)]
$100$ datasets consisting of $N=10000$ sample covariances, $2500$ for each cluster; in this case, we used the ``reduced`` algorithm described in Subsection \ref{subsec:large},  setting $N_{reduced}=200.$
\end{itemize}
In both cases, we assume that covariances are estimated using $n_i$ curves, with $n_i$ uniformly distributed in $\{5,\ldots, 10\}$. The curves are evaluated on a grid of 101 evenly spaced points.

\begin{figure}
    \centering
    \includegraphics[width=\textwidth]{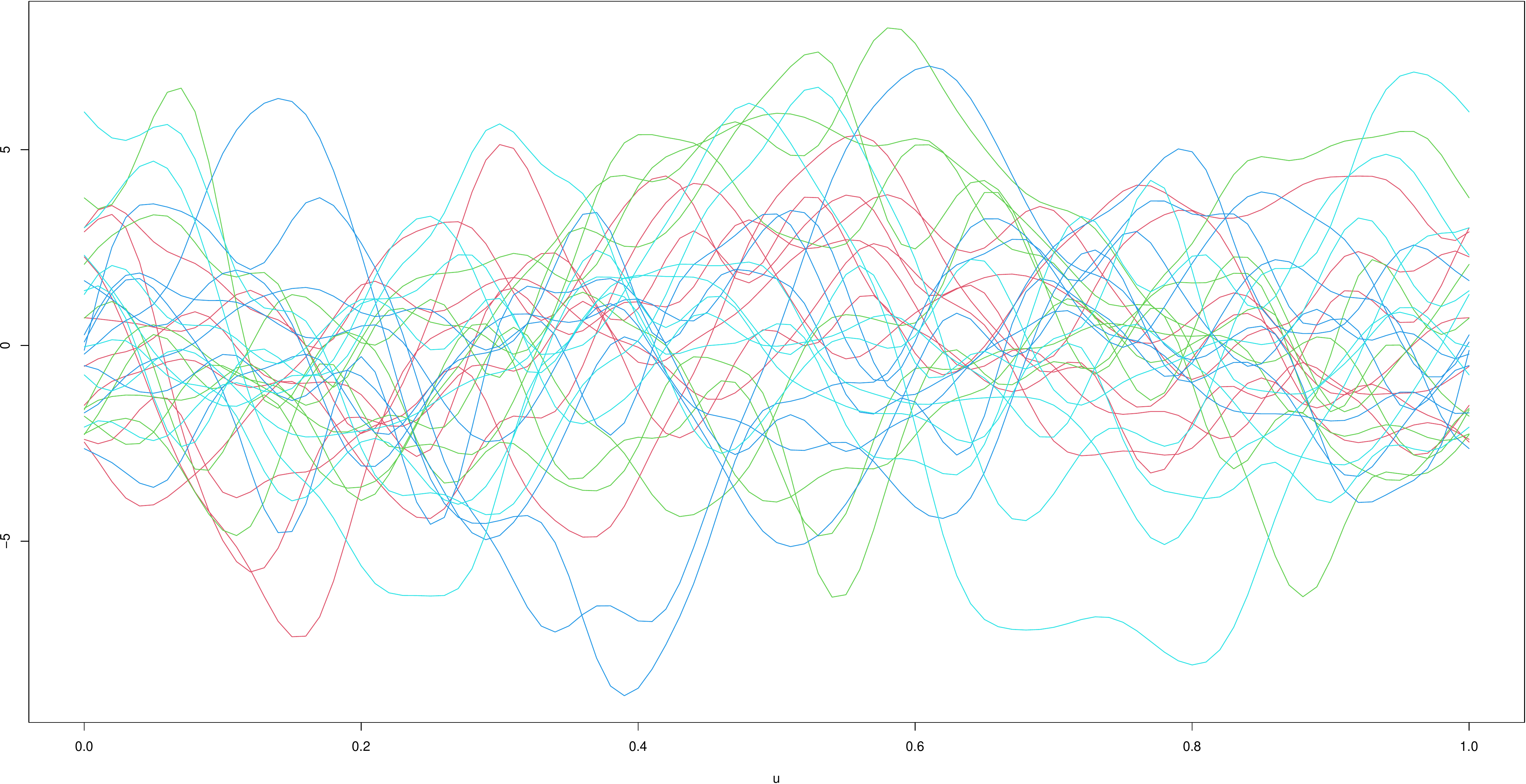}
    \caption{Examples of simulated functional data. The colours distinguish the four clusters.}
    \label{fig:X}
\end{figure}

\begin{figure}
    \centering
    \begin{tabular}{p{0.25\textwidth}p{0.25\textwidth}p{0.25\textwidth}p{0.25\textwidth}}
    \multicolumn{4}{c}{First cluster} \\
    \includegraphics[width=0.24\textwidth]{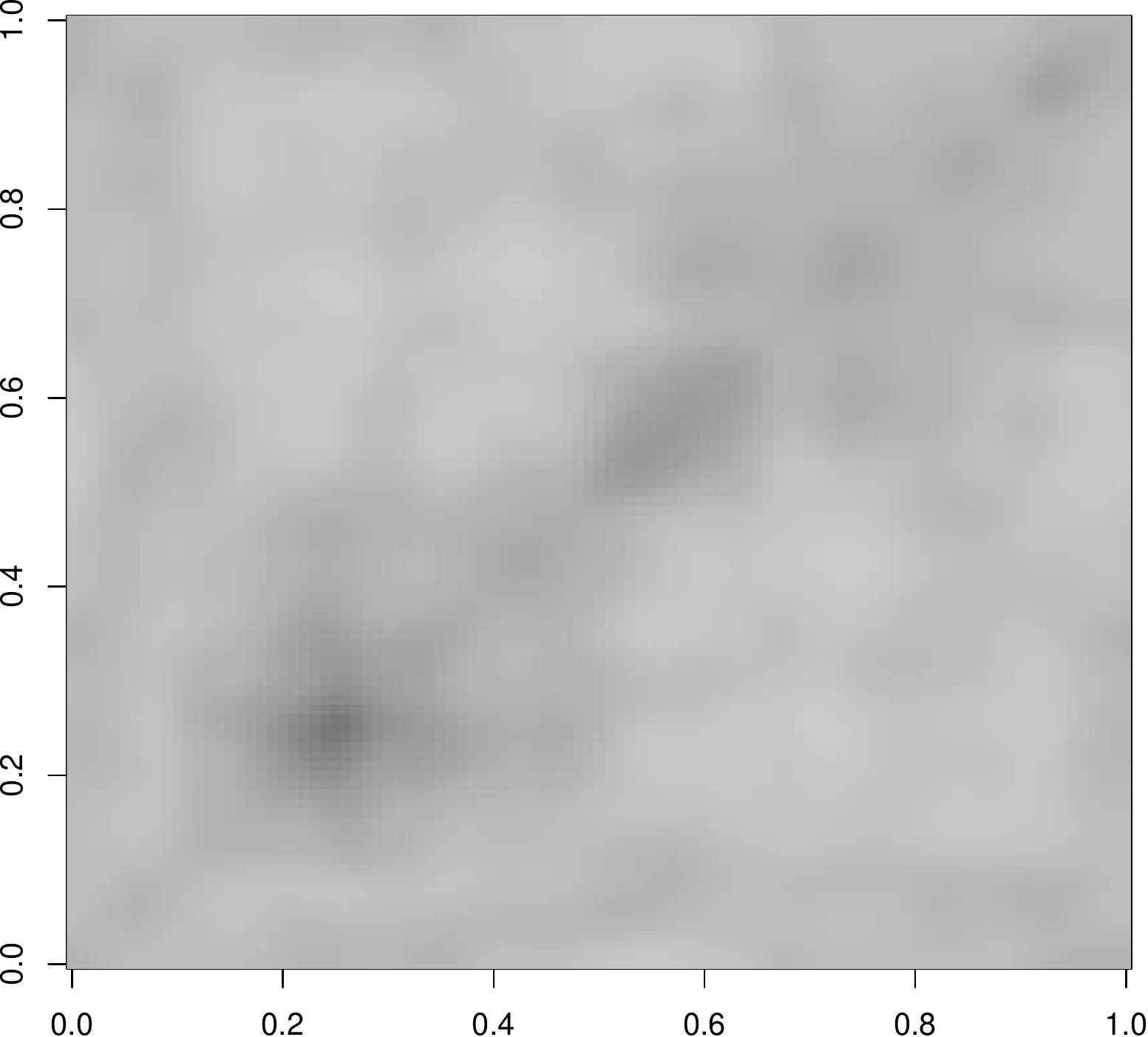}
    &
    \includegraphics[width=0.24\textwidth]{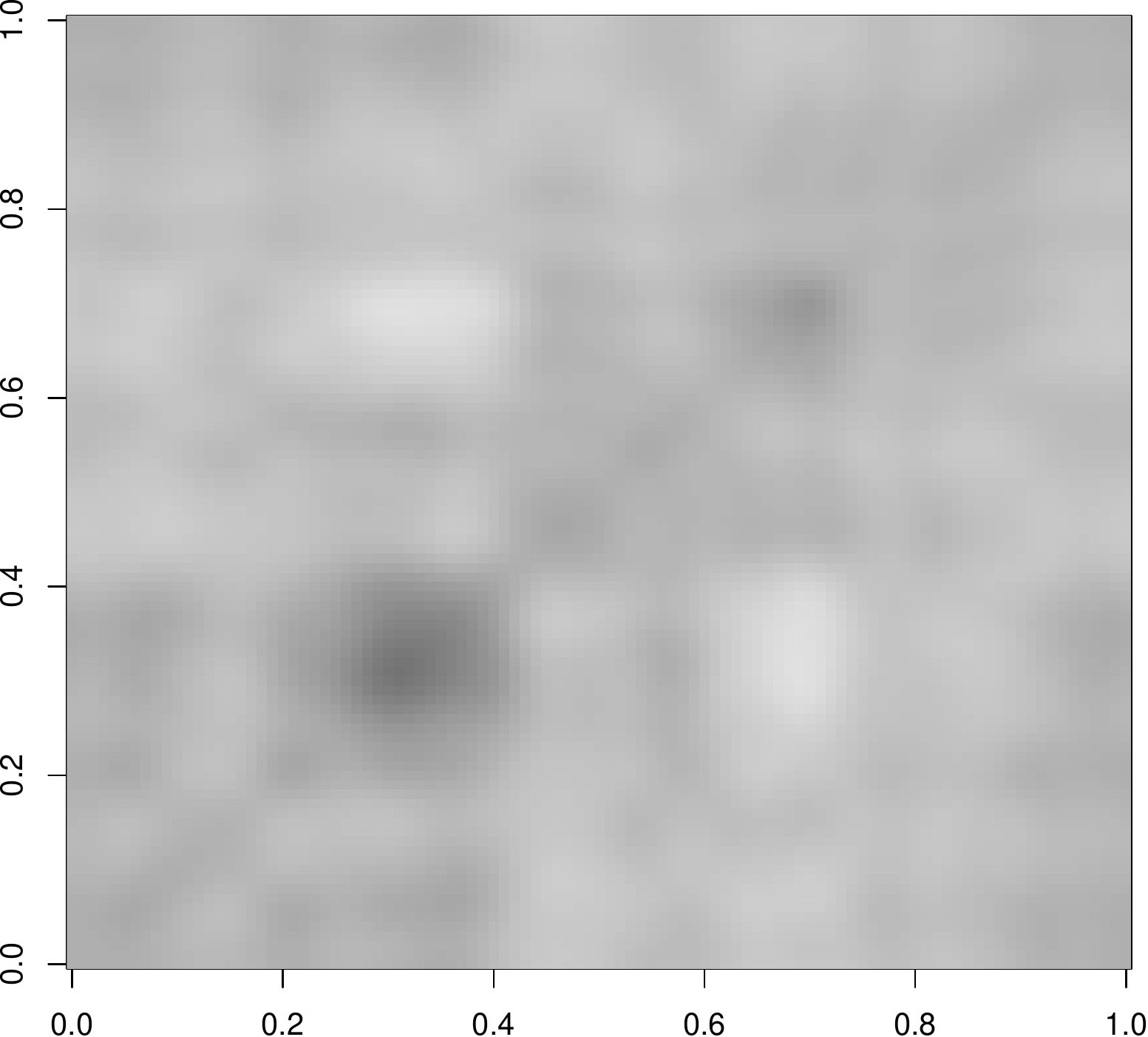}
    &
    \includegraphics[width=0.24\textwidth]{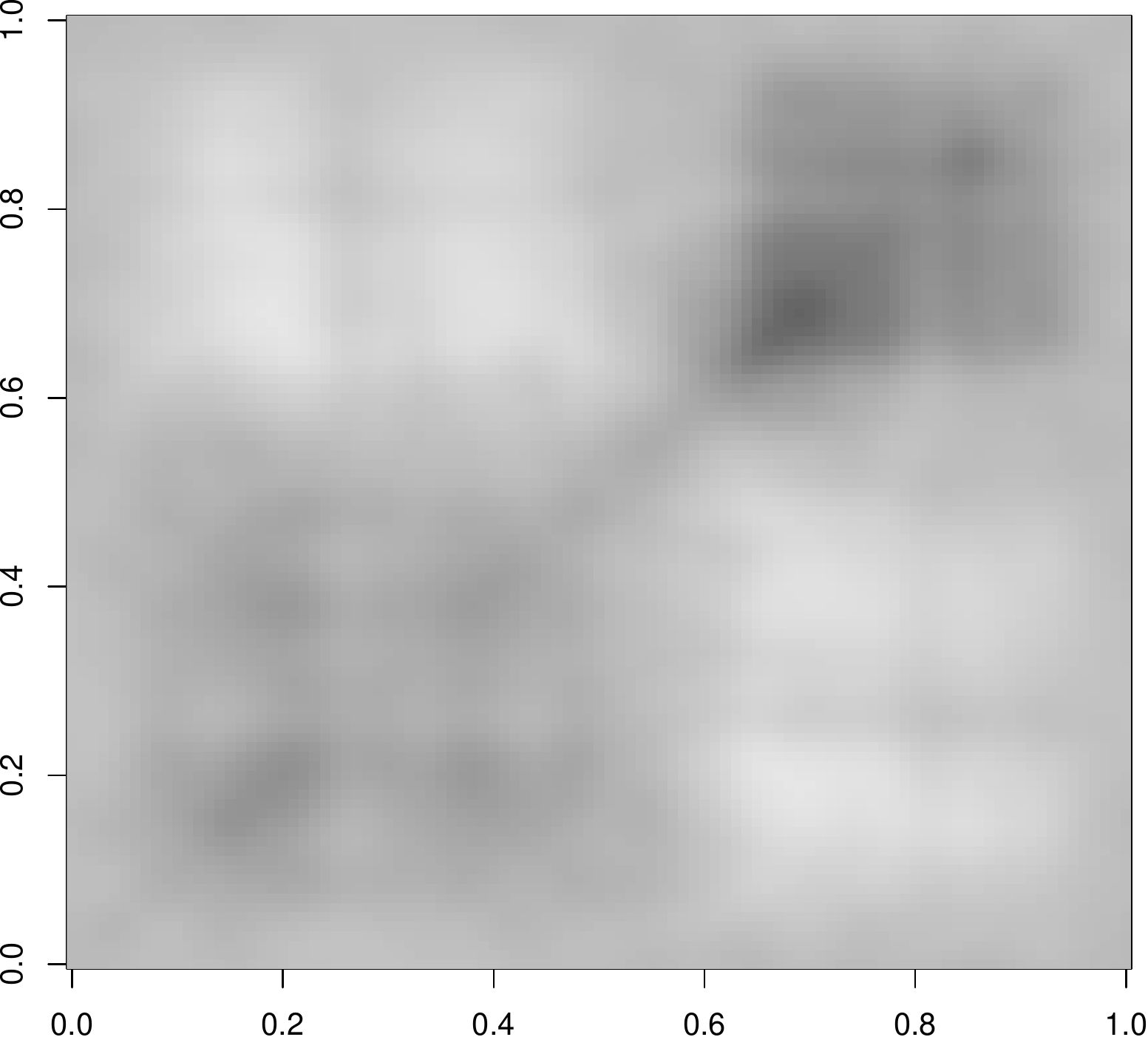}
    &
    \includegraphics[width=0.24\textwidth]{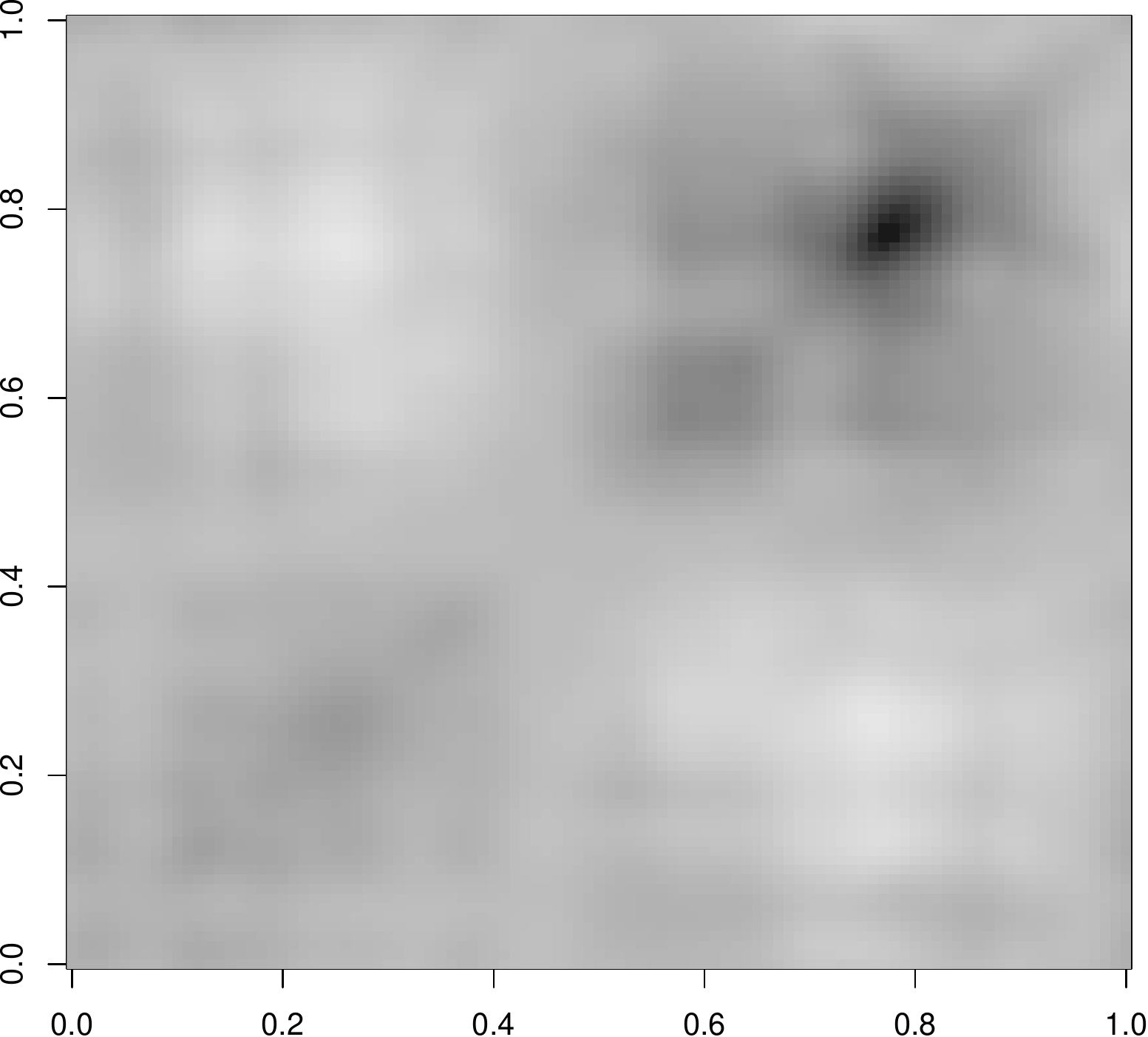}\\[10pt]
    \multicolumn{4}{c}{Second cluster} \\
    \includegraphics[width=0.24\textwidth]{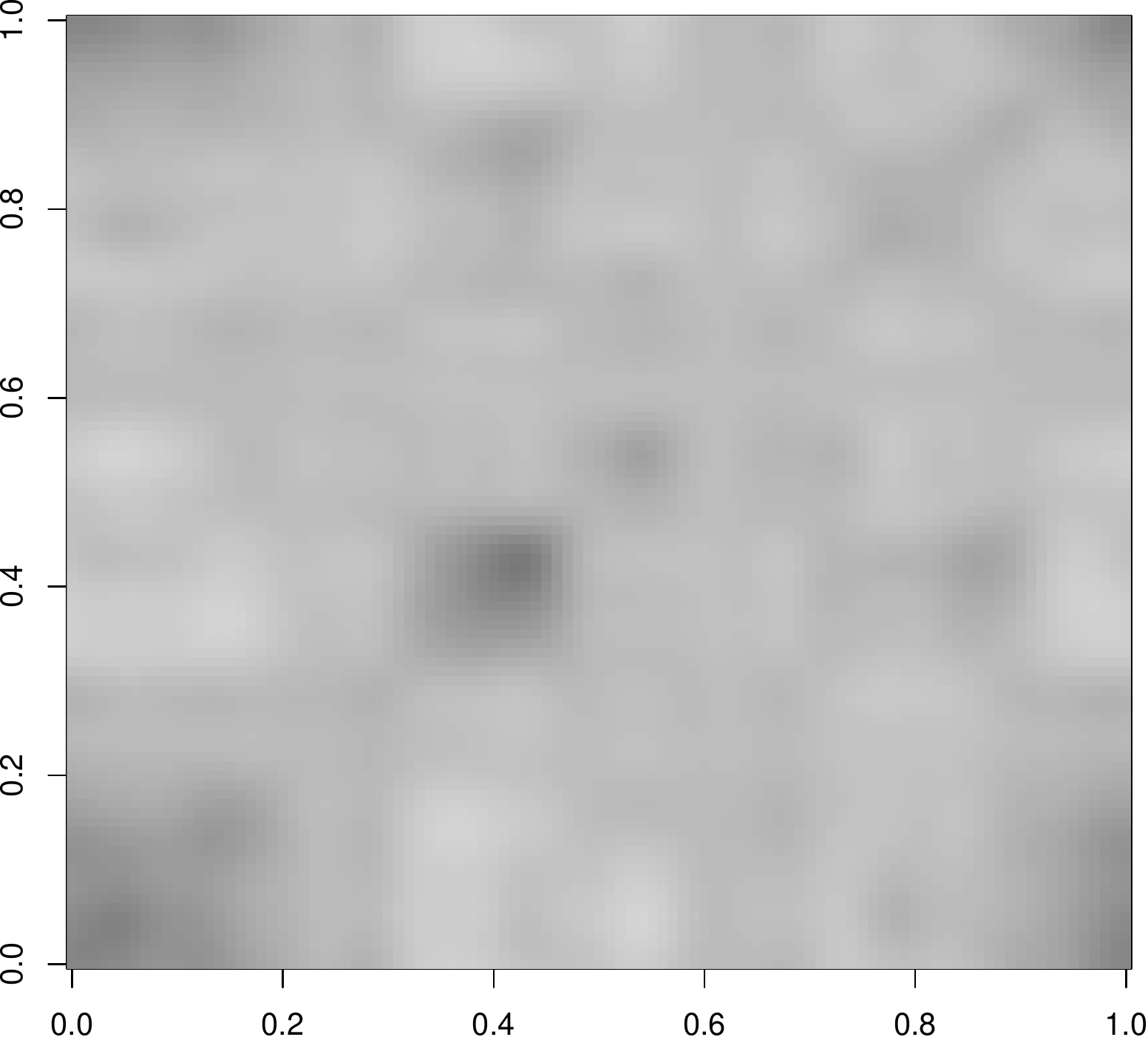}
    &
    \includegraphics[width=0.24\textwidth]{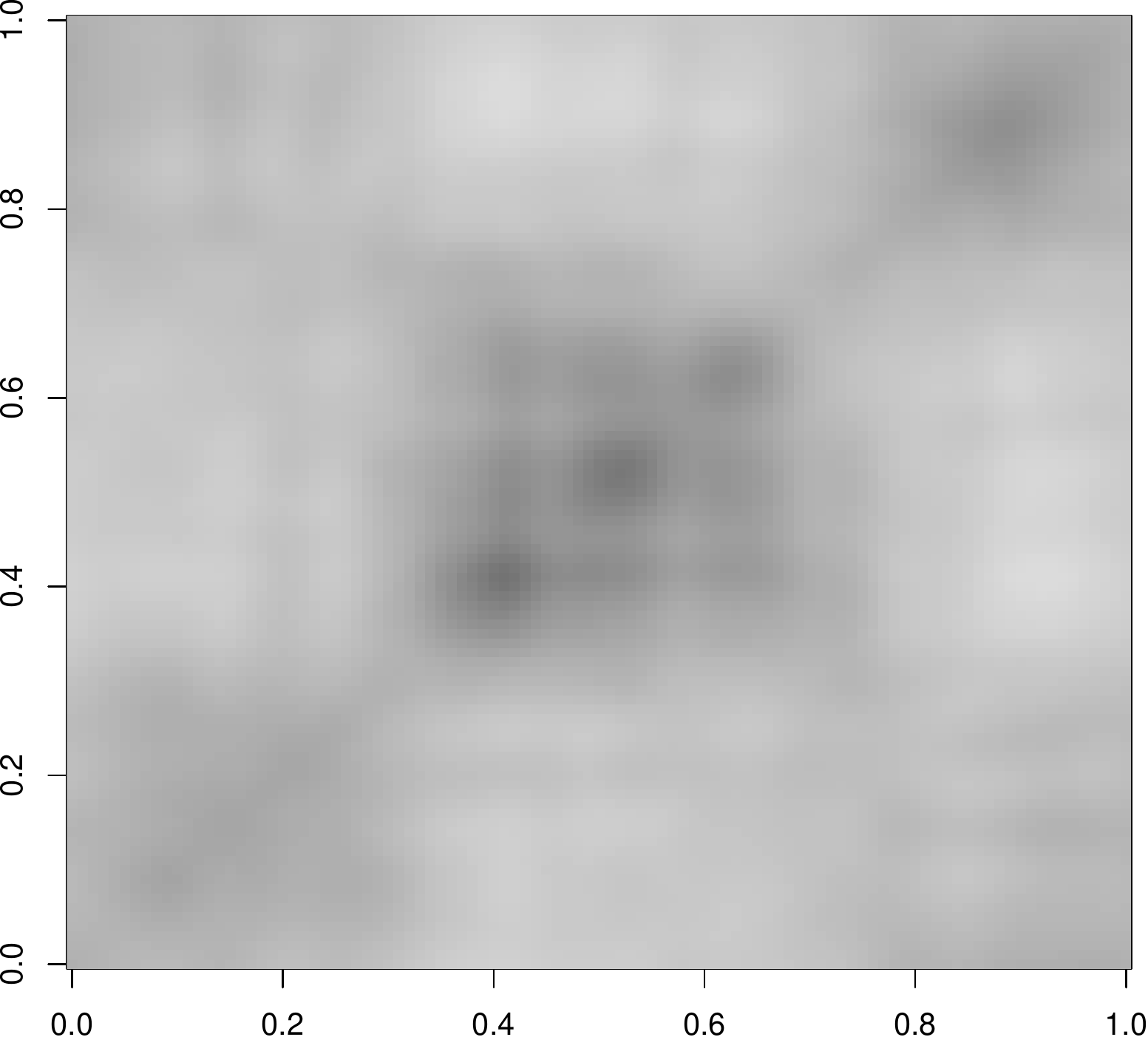}
    &
    \includegraphics[width=0.24\textwidth]{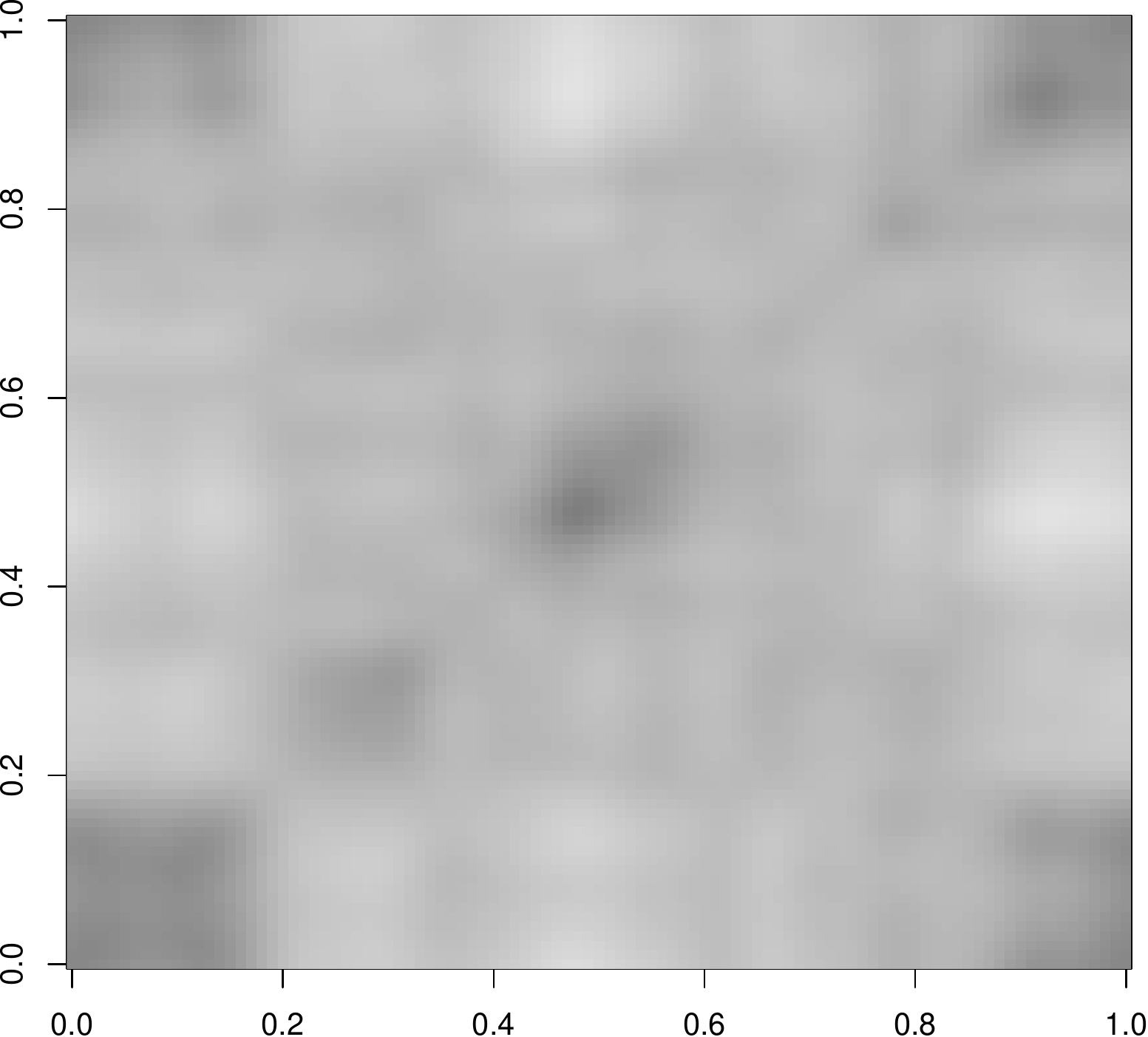}
    &
    \includegraphics[width=0.24\textwidth]{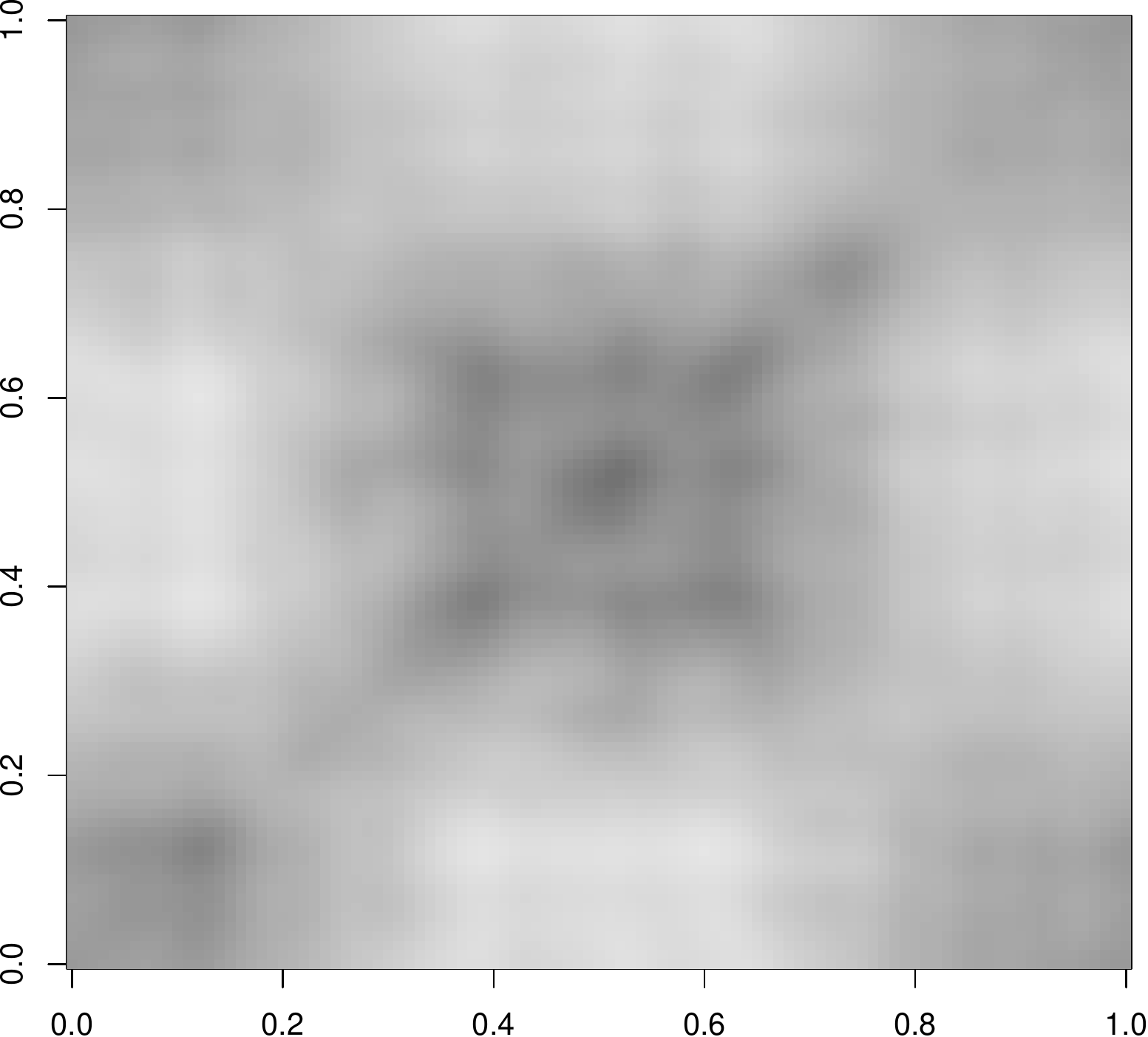}\\[10pt]
    \multicolumn{4}{c}{Third cluster} \\
    \includegraphics[width=0.24\textwidth]{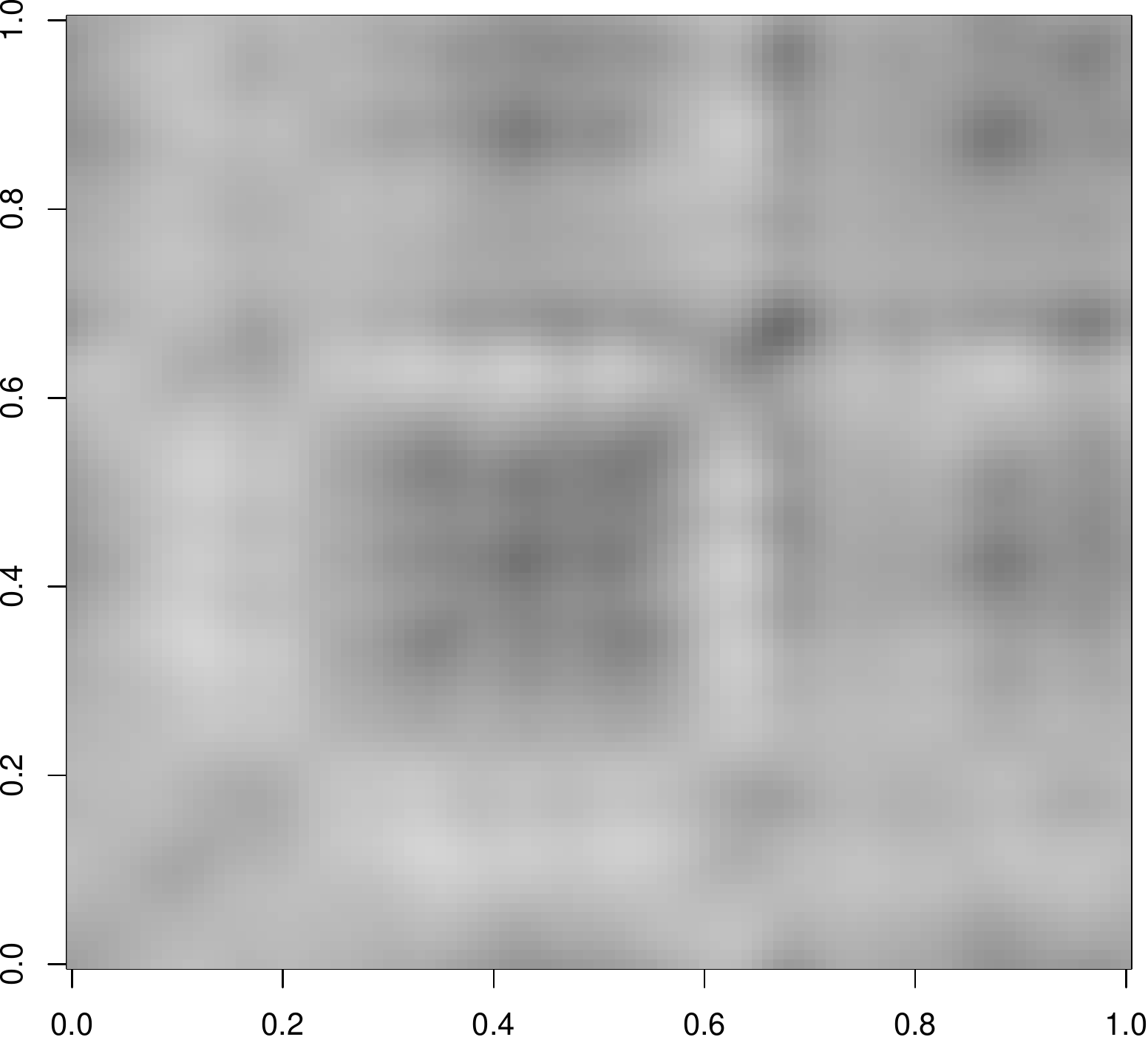}
    &
    \includegraphics[width=0.24\textwidth]{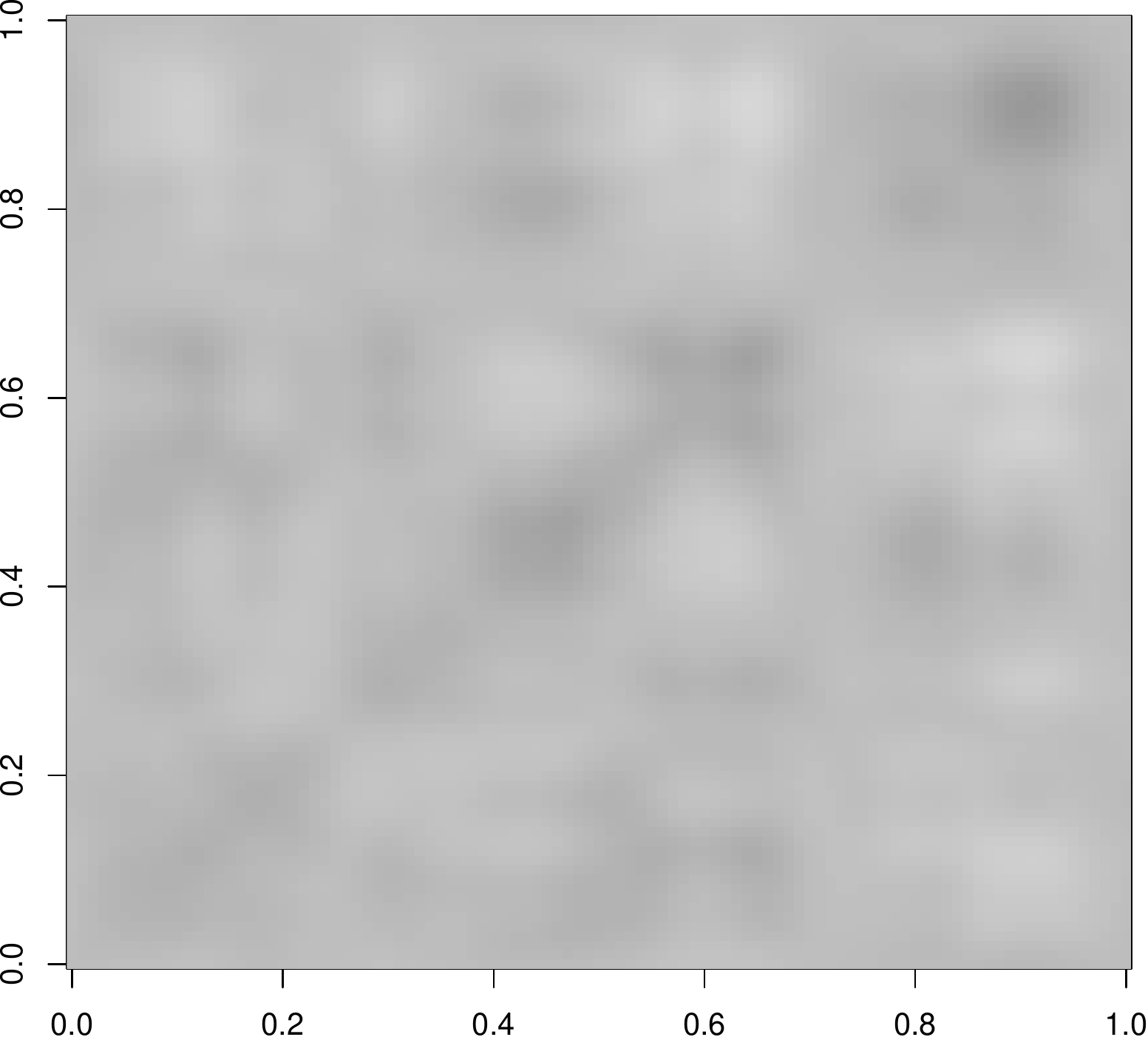}
    &
    \includegraphics[width=0.24\textwidth]{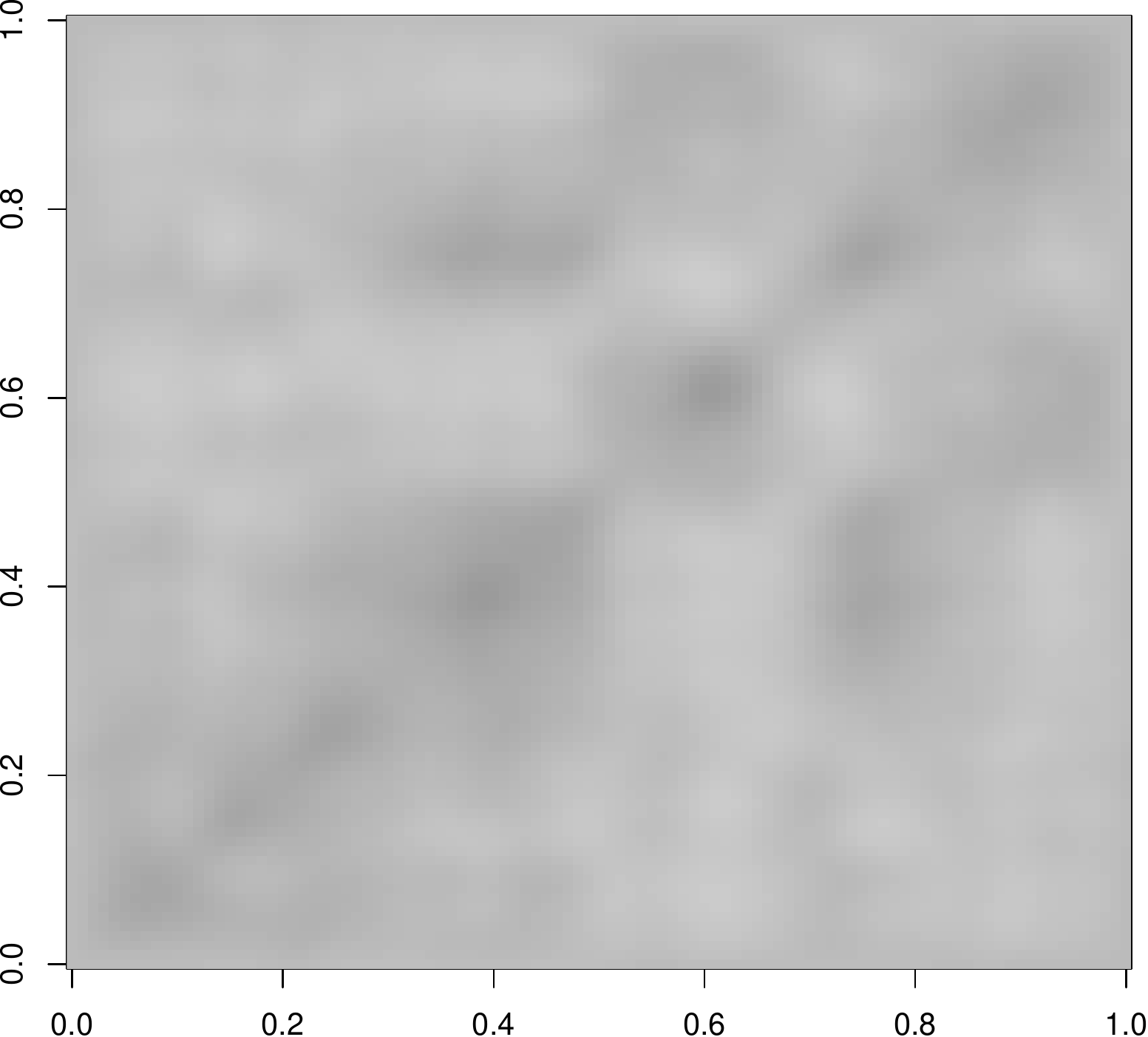}
    &
    \includegraphics[width=0.24\textwidth]{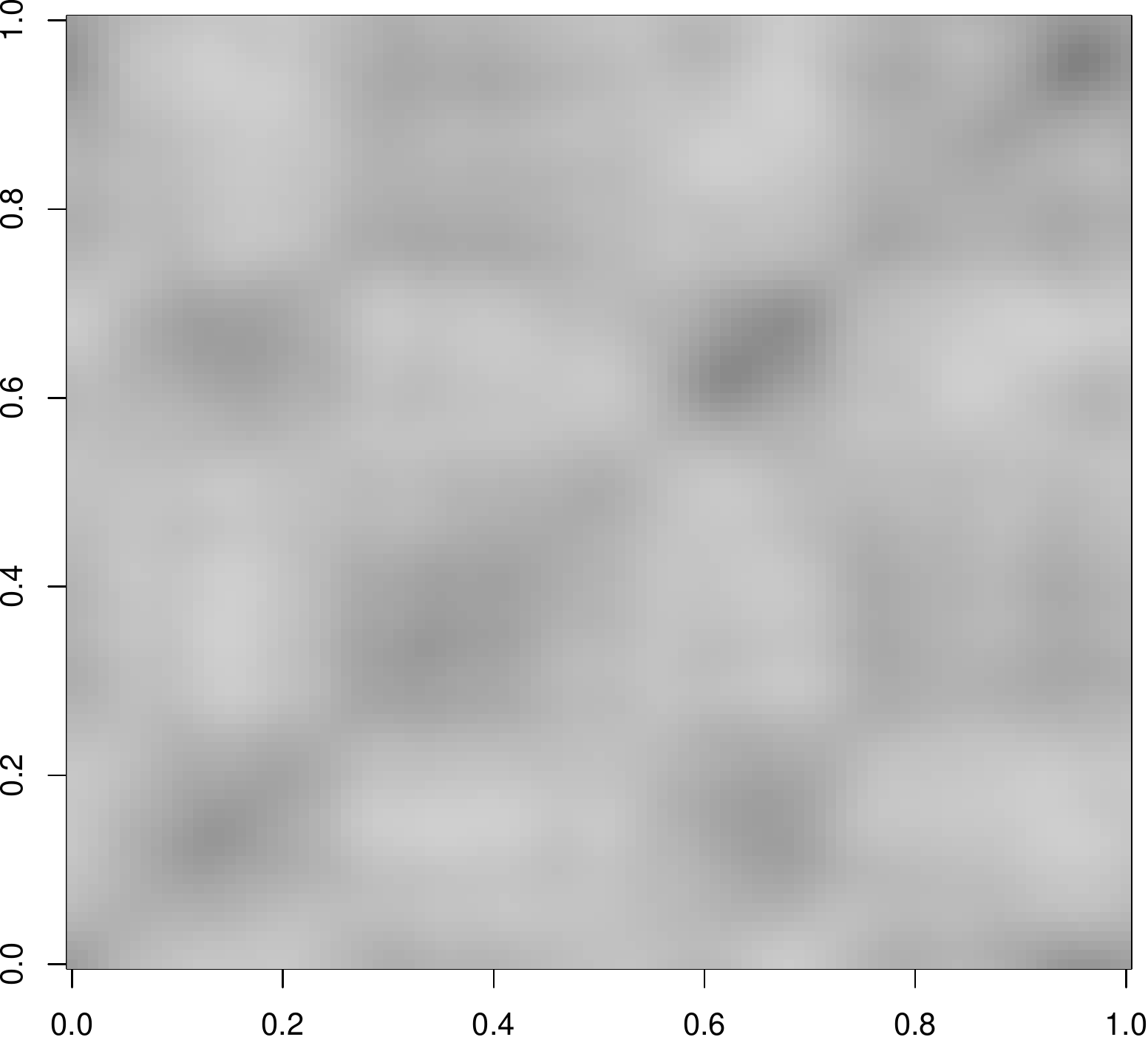}\\[10pt]
    \multicolumn{4}{c}{Fourth cluster} \\
    \includegraphics[width=0.24\textwidth]{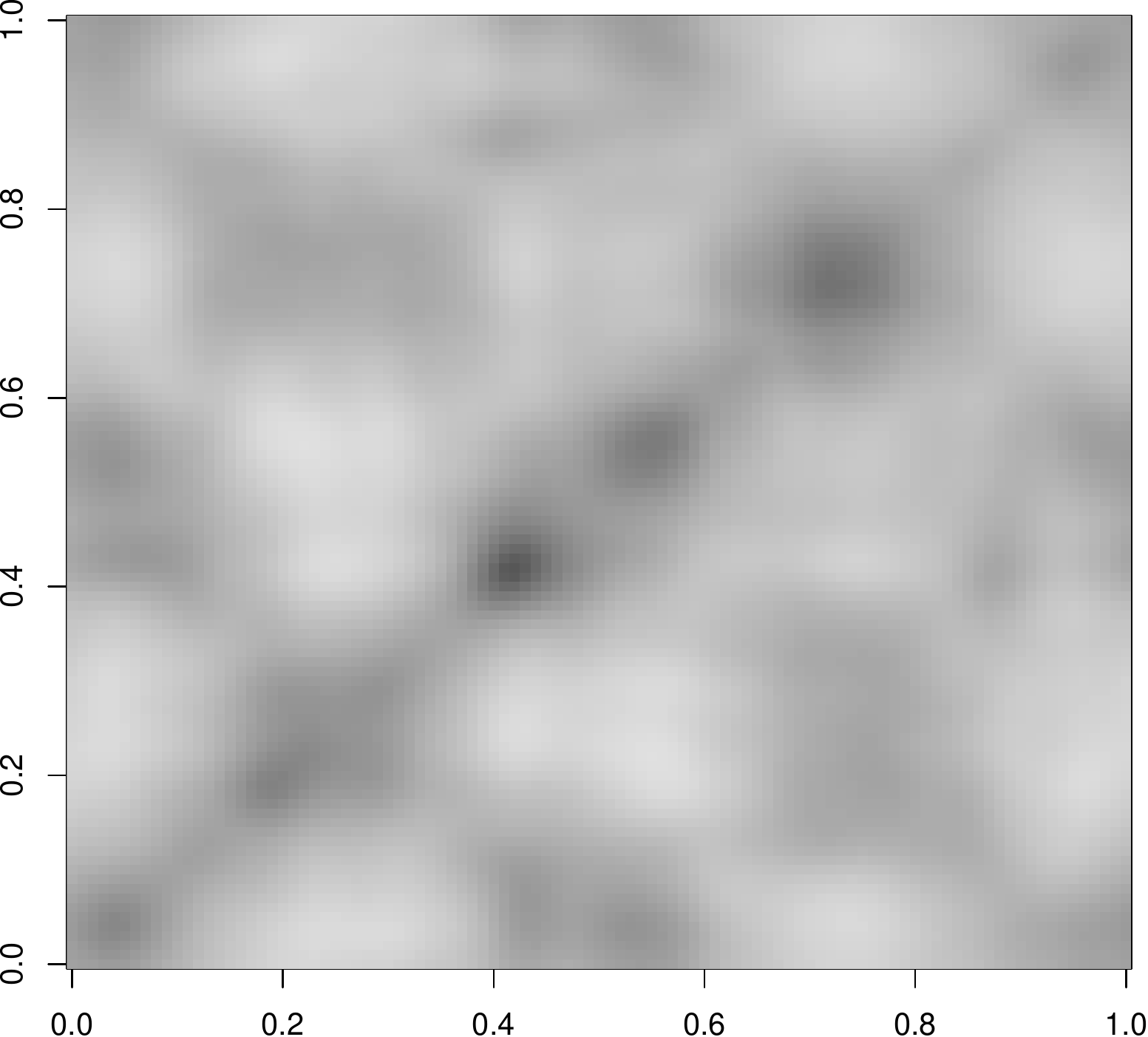}
    &
    \includegraphics[width=0.24\textwidth]{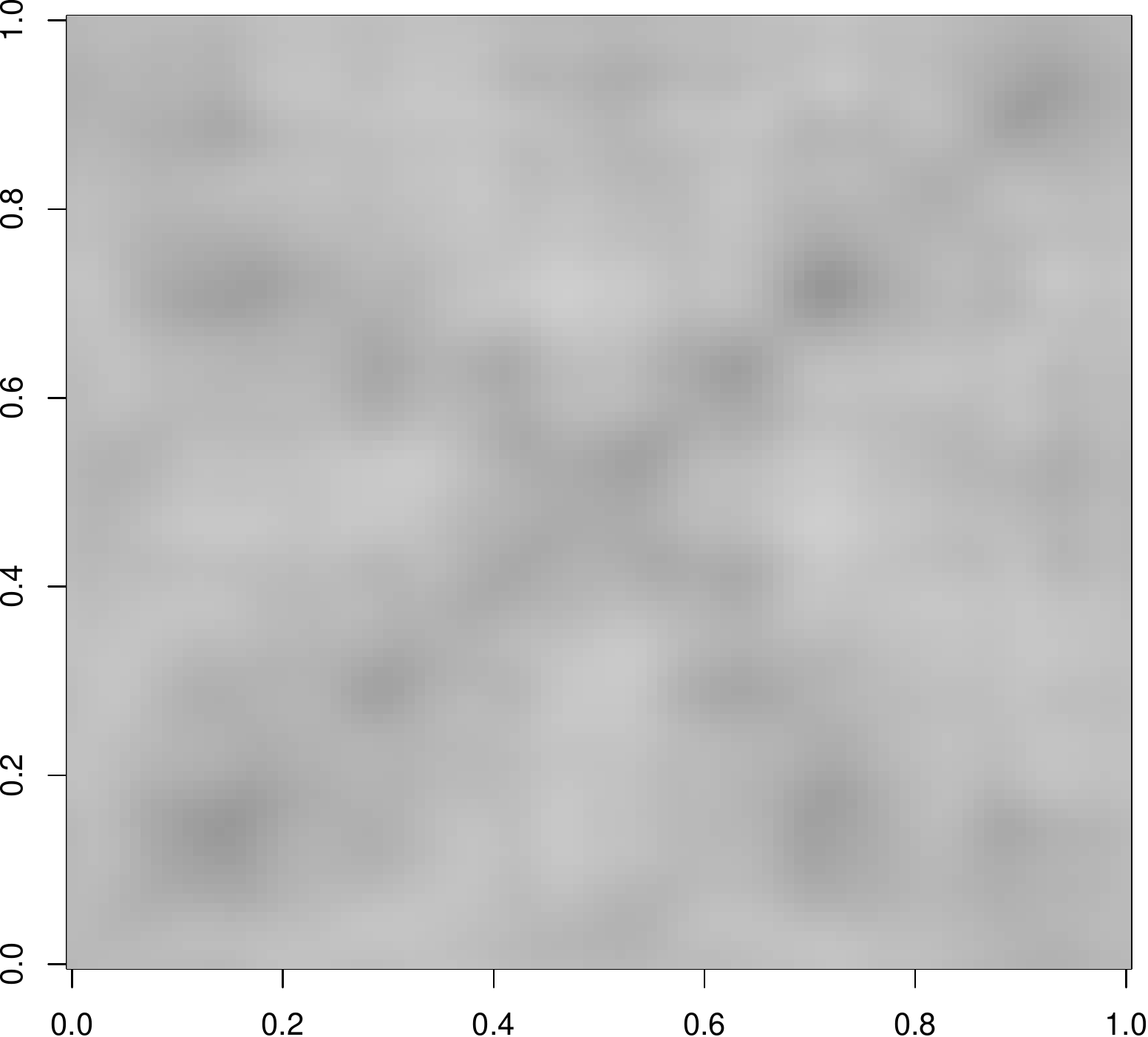}
    &
    \includegraphics[width=0.24\textwidth]{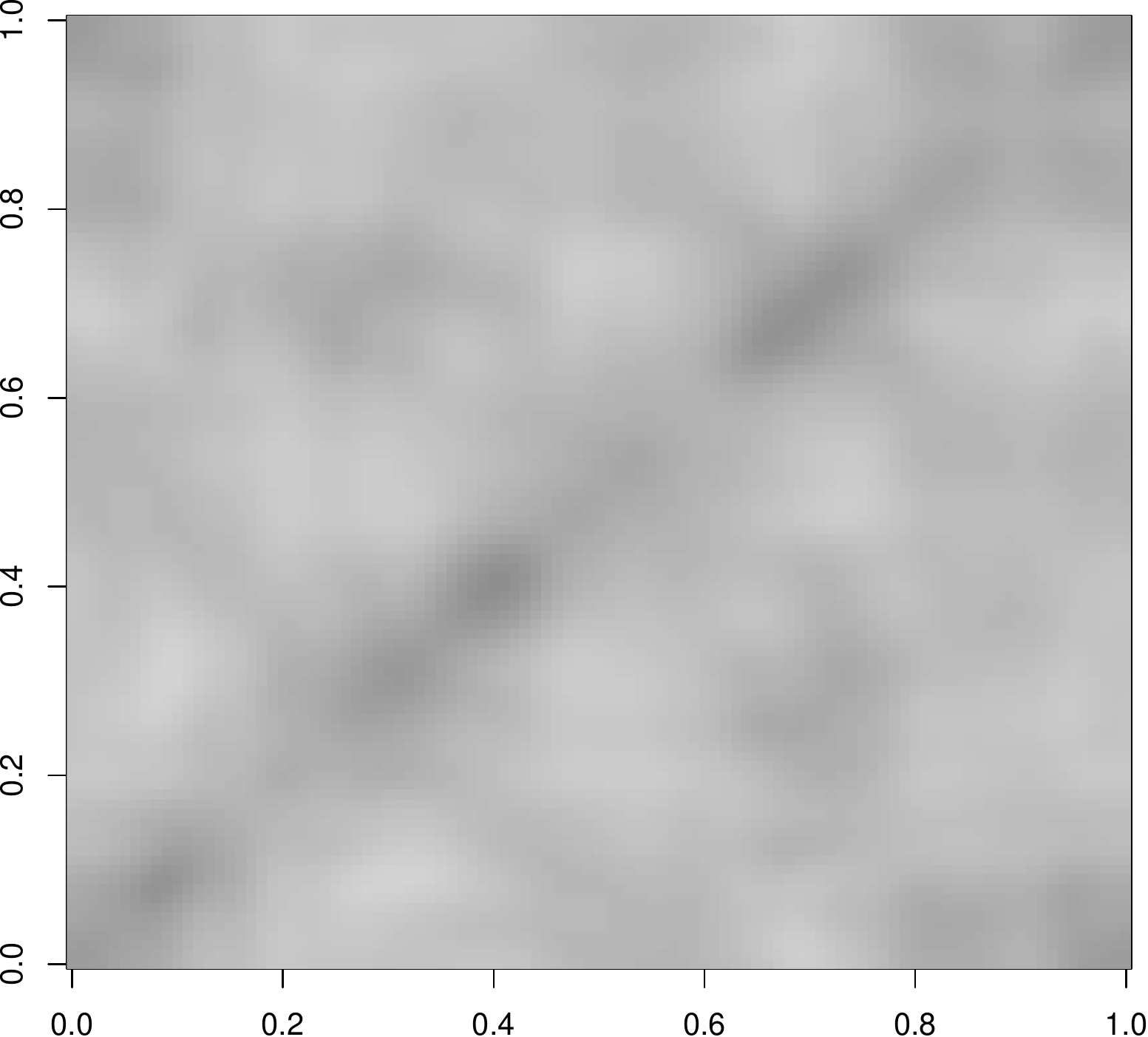}
    &
    \includegraphics[width=0.24\textwidth]{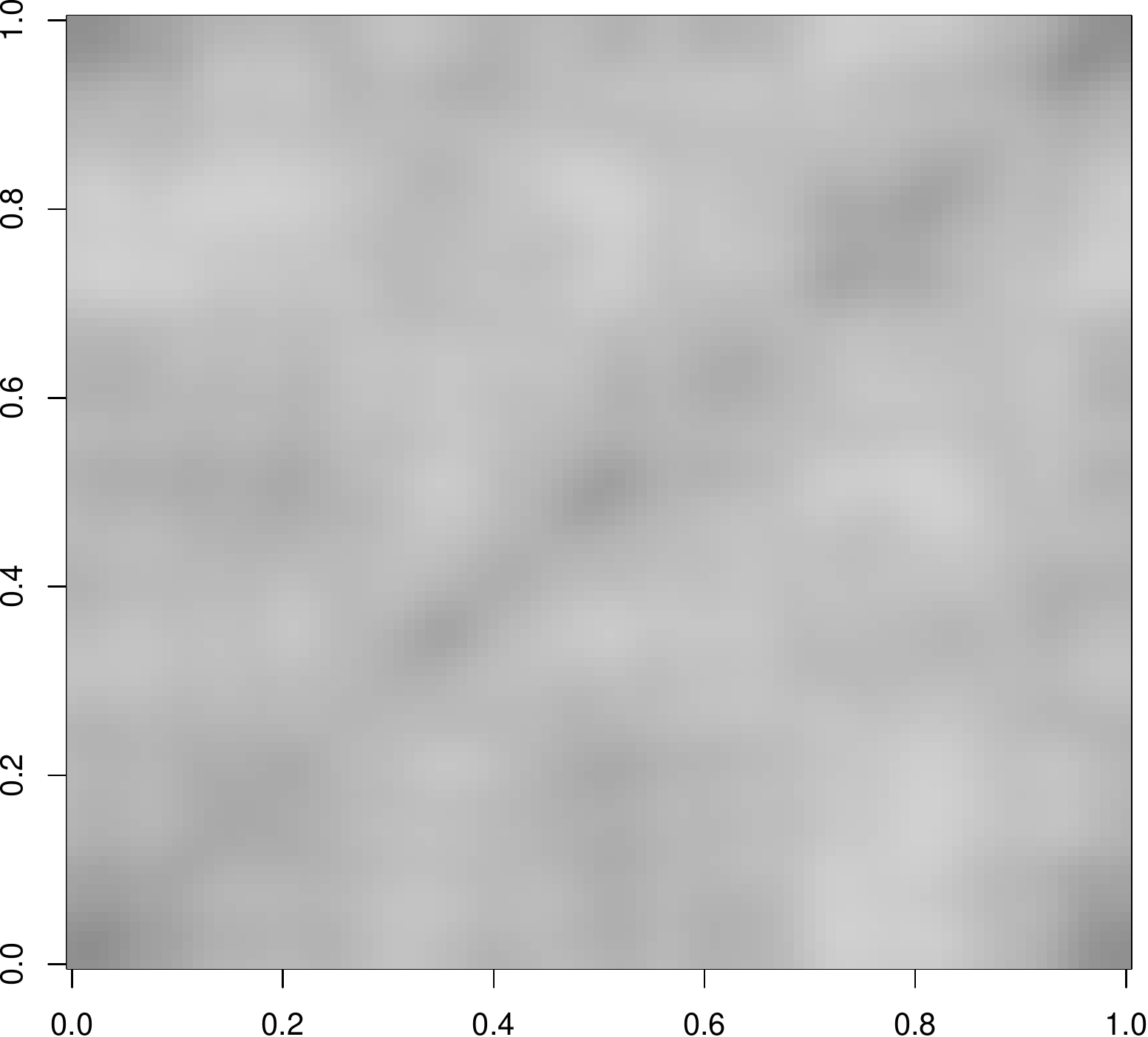}\\
    \end{tabular}
    \caption{Examples of sample covariance operators of the simulated functional data.}
    \label{fig:3syntheticChat}
\end{figure}

The differences between the four covariance operators characterising the clusters are not large. As $\sum_{r=0}^{32} \lambda^{2r} \approx 5$, the covariances essentially have 
five out of six ``variance components'' in common.  In addition, the sample sizes $n_i$ are small. 
As a consequence, it is difficult to recognize the four groups by visually inspecting the sample covariances
(or looking to the individual curves).
See the examples shown in Figures~\ref{fig:X} and \ref{fig:3syntheticChat}.

\begin{figure}
    \centering
    \includegraphics[width=\textwidth, height=0.7\textwidth]{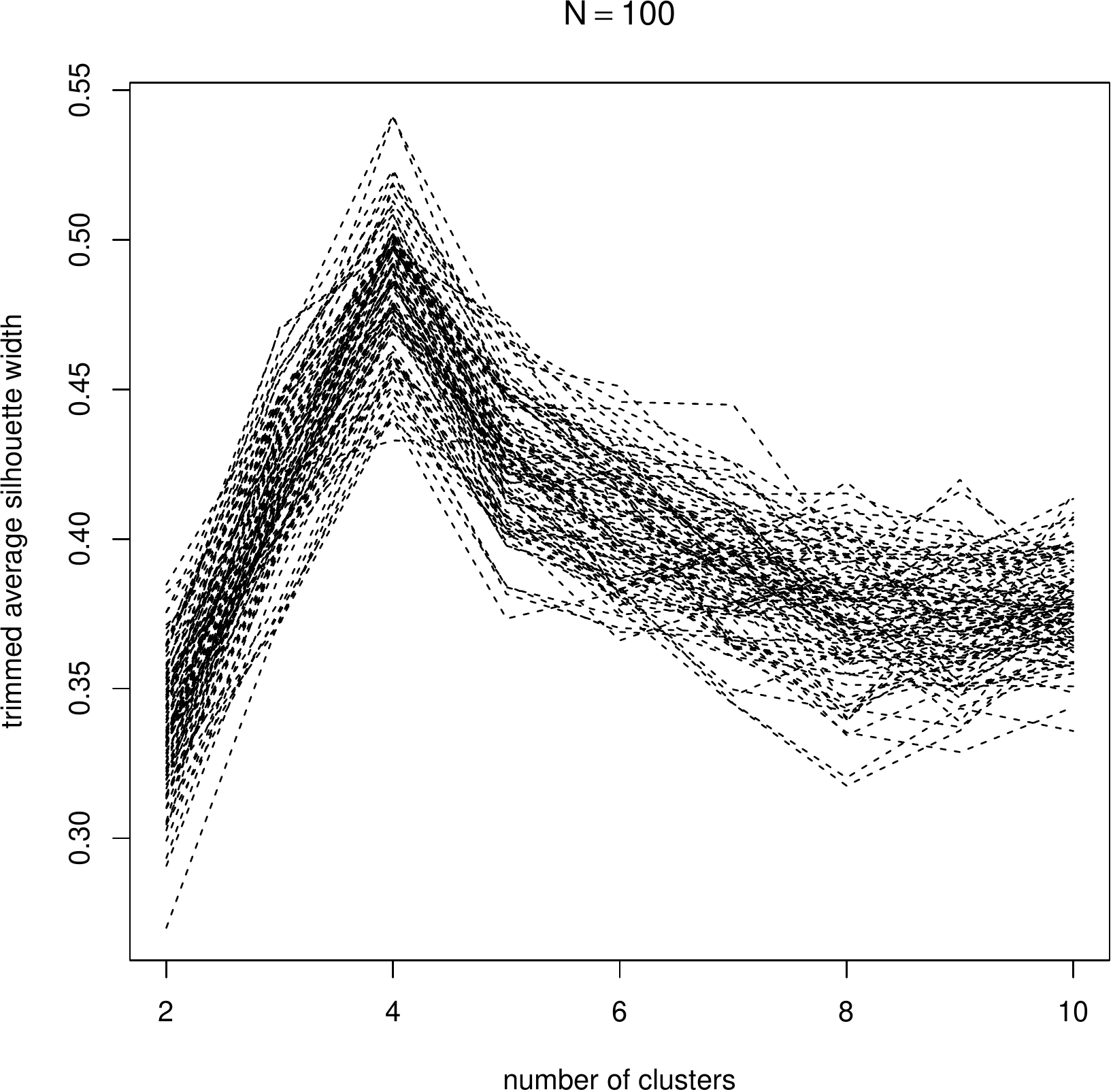}\\[15pt]
    \includegraphics[width=\textwidth, height=0.7\textwidth]{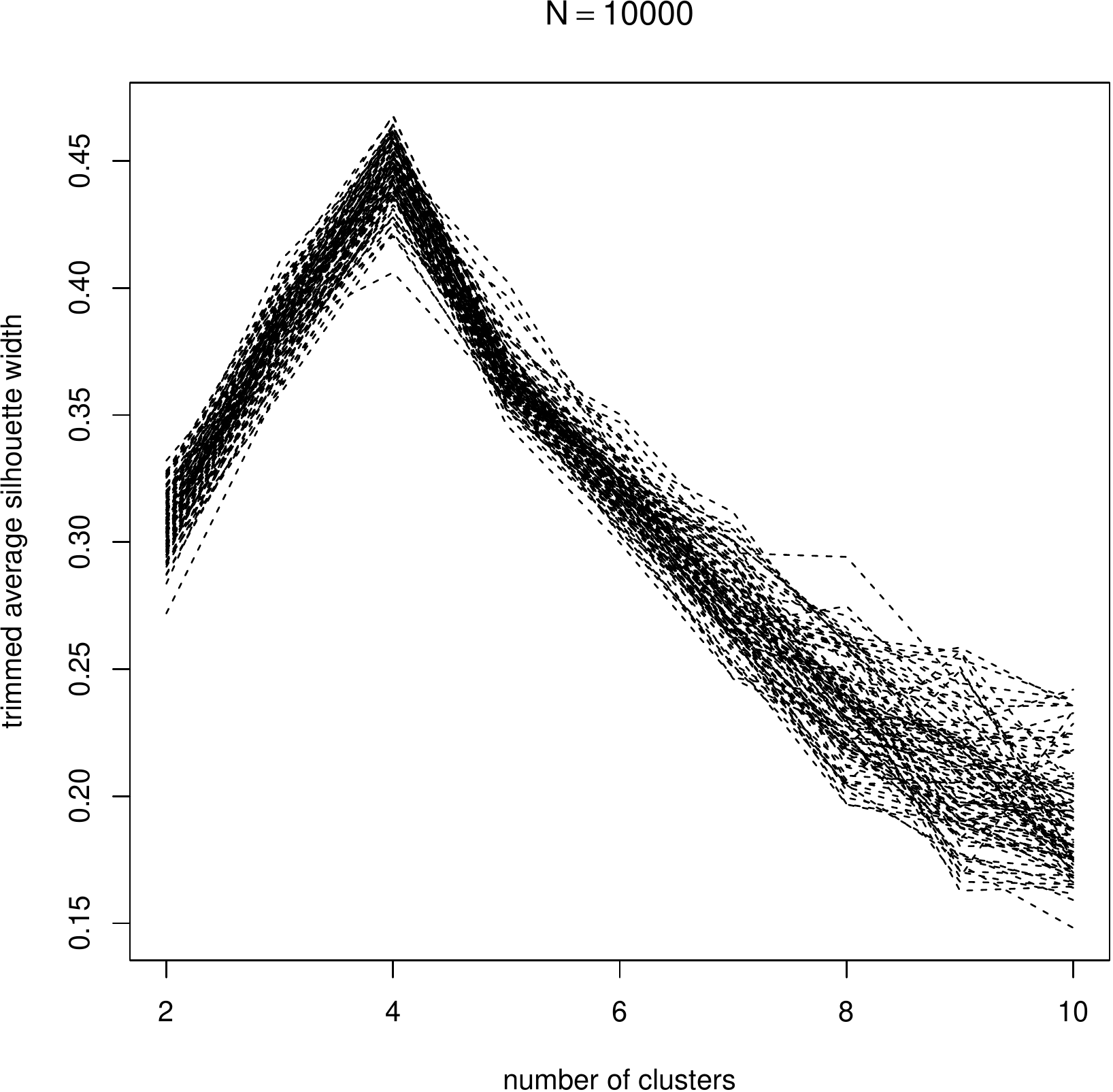}\\
    \caption{Synthetic data: The trimmed average silhouette width for $100$ simulated datasets with $N=100$ and $N=10000$ sample covariances. The true number of clusters is $K=4$.}
    \label{fig:3synthetic-tasw}
\end{figure}

For the two considered cases ($N=100$ and $N=10000$), Figure~\ref{fig:3synthetic-tasw} displays the profiles of the trimmed average silhouette width obtained by applying the cluster algorithm for $K=2,\ldots,10$ to each of the simulated datasets. 
The  profiles point to the correct number of clusters. In all the cases, the maximum was obtained exactly at $K=4$. Of course, in other scenarios, the performance of the suggested index can be inferior to that observed here. 

\begin{figure}
    \centering
    \includegraphics[width=\textwidth, height=0.8\textwidth]{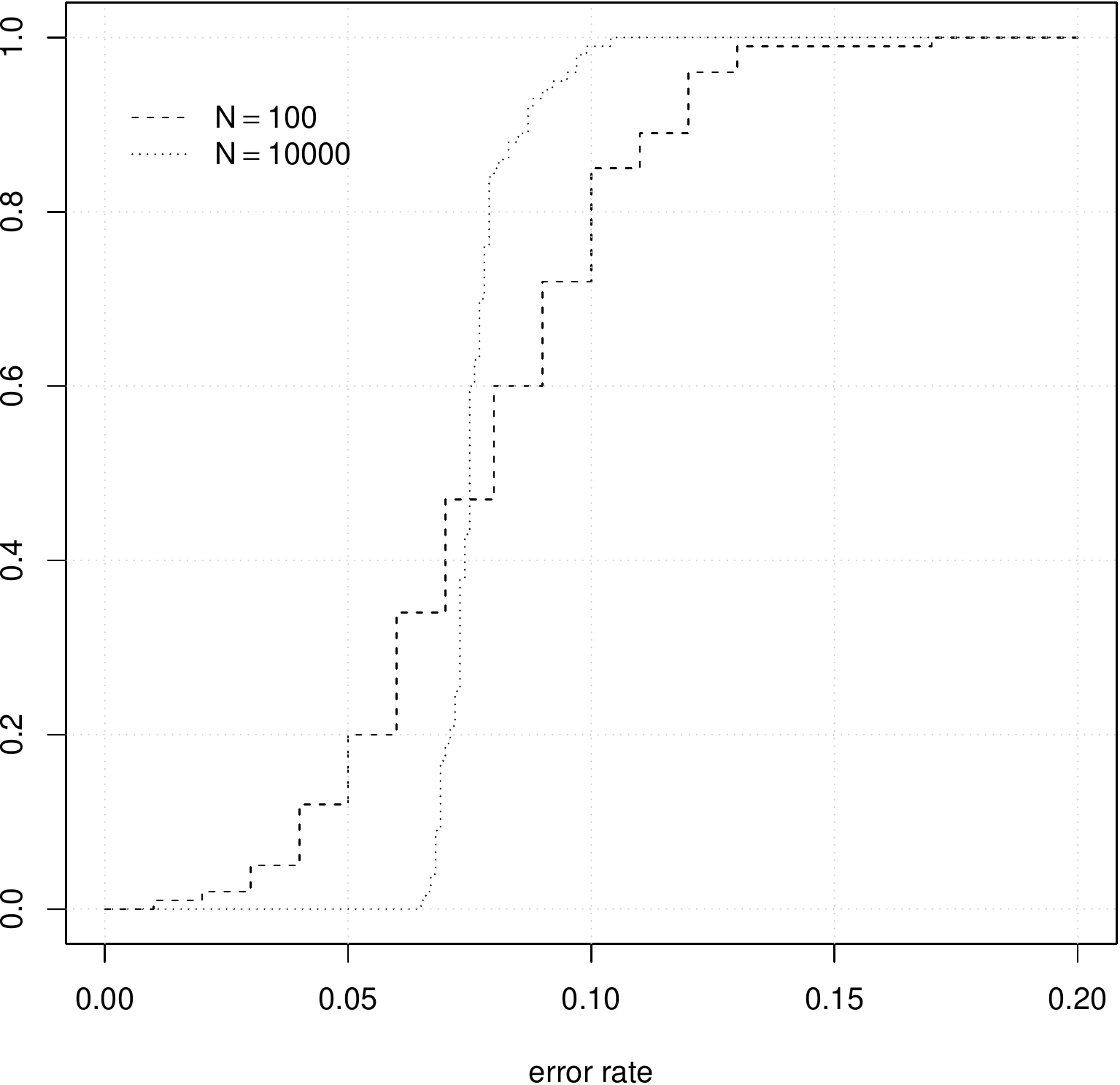}
    \caption{Synthetic data: cumulative distribution function of the error proportion of the ``nearest allocation rule'' in $100$ simulated datasets.}
    \label{fig:3synthetic-errs}
\end{figure}

In the remainder of this subsection, we focus on the quality of the classification results obtained for $K=4$.
Figure~\ref{fig:3synthetic-errs} displays the empirical cumulative distribution functions of the error rates 
observed in the different simulated datasets  when the ``nearest allocation rule'' 
\begin{center}
    \itshape classify $\widehat\Sigma_i$ in the group whose prototype is $\overline\Sigma_{i_a}$ if
    $i_a =\arg \max_j \pi_{i,j} = \arg \min_r \Pi\bigl(\widehat\Sigma_i, \overline\Sigma_j\bigr)$
\end{center}
is used. When $N=100$ and $N=10000$, the figure shows that 
the median of the classification errors is around 
$7\%$. In addition, observe that the error rates show little variability in the $N=10000$ case, which points to the limited effects of the additional randomization due to the selection of the $N_{reduced}=200$ covariances.
Table~\ref{tab:3nearest-errors} shows that the membership grades
$\pi_{i,j}$ can capture the uncertainty of the classification, and so, in some sense, 
\emph{to predict the errors}. 
In particular, the table shows the classification error rates of the ``nearest allocation rule'' subdivided for the
levels of the credibility coefficients introduced in \eqref{eqn:credibility}. Observe that the error rates
are between $50\%$ and $60\%$ when the credibility of the allocation is low, that is, for about $5\%$ of the sample covariances
for which $\max_j \pi_{i,j}$ was less than $0.5$. Then, the error rate decreases as the credibility of the allocation increases,  reaching a value close to zero  for about $70\%$ of the sample covariances for which $\max_j \pi_{i,j} > 0.9$. It is also interesting to observe that the error rates are comparable, if not slightly lower,  
when $N=10000$ than when $N=100$. This finding confirms the limited impact of estimating the
cluster barycenters using only part of the available sample covariances.

\begin{table}[t]
\centering
\begin{tabular}{lcccccc}
\hline
  & \multicolumn{6}{c}{Credibility of the ``nearest allocation rule''} \\
  & $[0,0.5]$ & $(0.5,0.6]$ & $(0.6,0.7]$ & $(0.7,0.8]$ & $(0.8,0.9]$ & $(0.9,1]$\\
\hline
  \\[-6pt]
  & \multicolumn{6}{c}{$N=100$} \\
$\%$ covariances & 0.047 & 0.049 & 0.051 & 0.060 & 0.100 & 0.693\\
Error rate & 0.571 & 0.425 & 0.278 & 0.147 & 0.054 & 0.003\\
\\[-6pt]
  & \multicolumn{6}{c}{$N=10000$} \\
$\%$ covariances        &   0.048 &     0.047 &      0.050 &     0.063 &     0.104 &   0.686 \\
Error rate &   0.552 &     0.394 &      0.260 &     0.142 &     0.056 &   0.004\\
\hline
\end{tabular}
\caption{Error rates of the ``nearest allocation rule'' for different levels of the credibility 
of the allocation. The credibility is given by $C_{i}=\max_{j=1,\ldots,K} \pi_{i,j}^K$.}
\label{tab:3nearest-errors}
\end{table}

\subsection{Near-infrared gasoline spectra}

In this subsection, we consider the gasoline near-infrared (NIR) spectra collected at an oil refinery and analyzed by \citet{capizzimasarotto_2018}. Production engineers  collected 12 gasoline samples per day for 47 consecutive days. For each sample,  the NIR absorbance is available for wavelengths from 900 to 1700~nm in intervals of 2~nm (thus, each spectrum is observed at 401 different wavelengths). 
\begin{figure}[t]
    \centering
    \includegraphics[width=\textwidth]{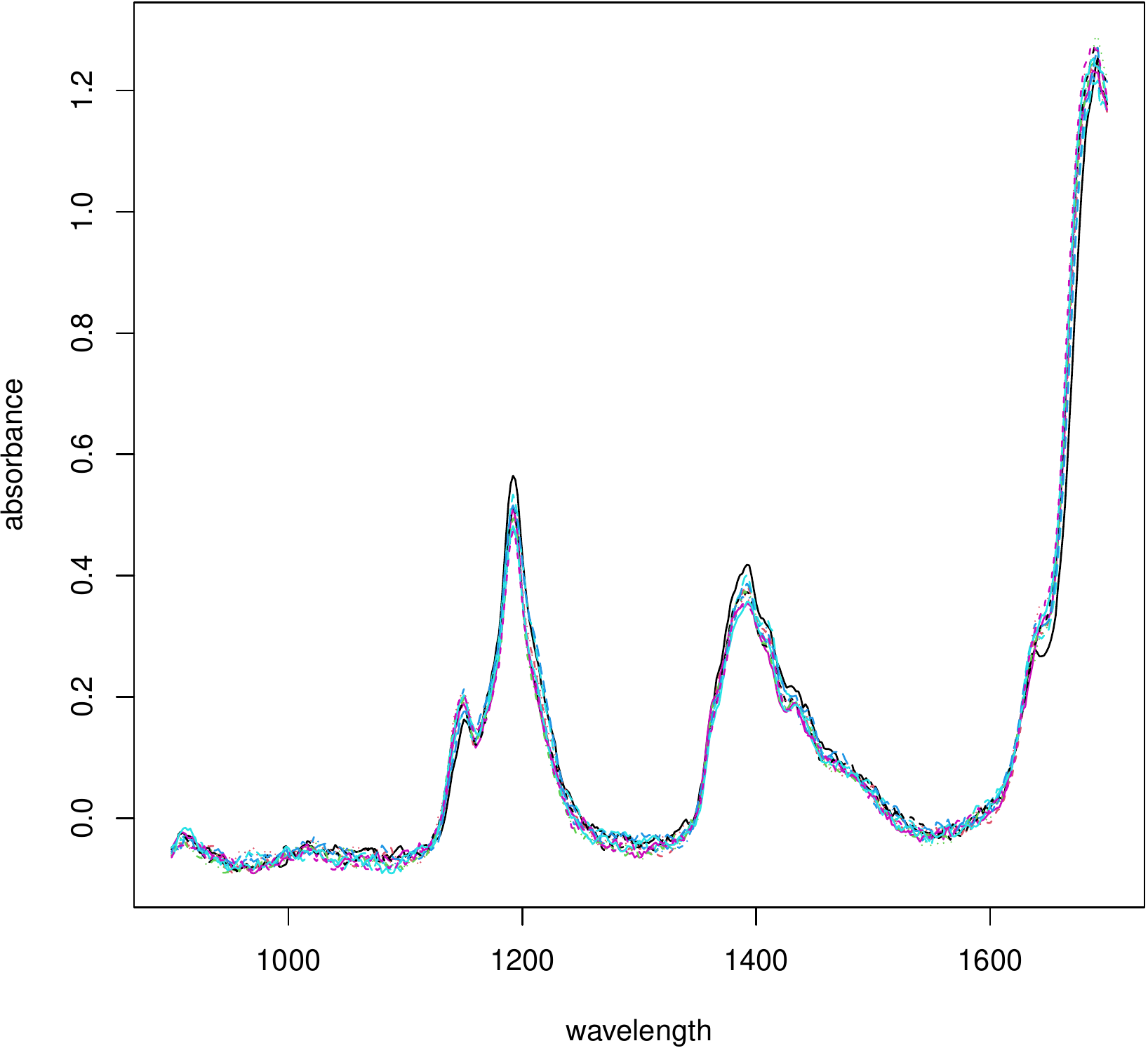}
    \caption{NIR spectra: the 564 observed curves.}
    \label{fig:spectra}
\end{figure}
Figure \ref{fig:spectra}
shows the $12 \times 47 = 564$ spectra. The data were gathered to understand if during the study there was day-to-day variability in the production conditions. However, the 12 samples collected each day were gathered several minutes apart, and therefore, it can be assumed that they were produced under the same set of conditions. Therefore, similar to what \citet{capizzimasarotto_2018} did,
we assume that the data consist of $N=47$ samples (one for each day), each comprising $n=12$ curves.
\citet{capizzimasarotto_2018} applied univariate distribution-free control charts separately to the first principal components (PC), and found an increase in the variability of the second, third,  and fourth PCs during the days from 32 to 37.  No differences in location in any PC, or in the variability of the other PCs, was identified.  The instability in days 32--37 was then confirmed by the production engineers and attributed to a transitory malfunction of the automatic process adjustments that resulted in an increase in the variability of the product characteristics. 

\begin{figure}[t]
    \centering
    \includegraphics[width=\textwidth]{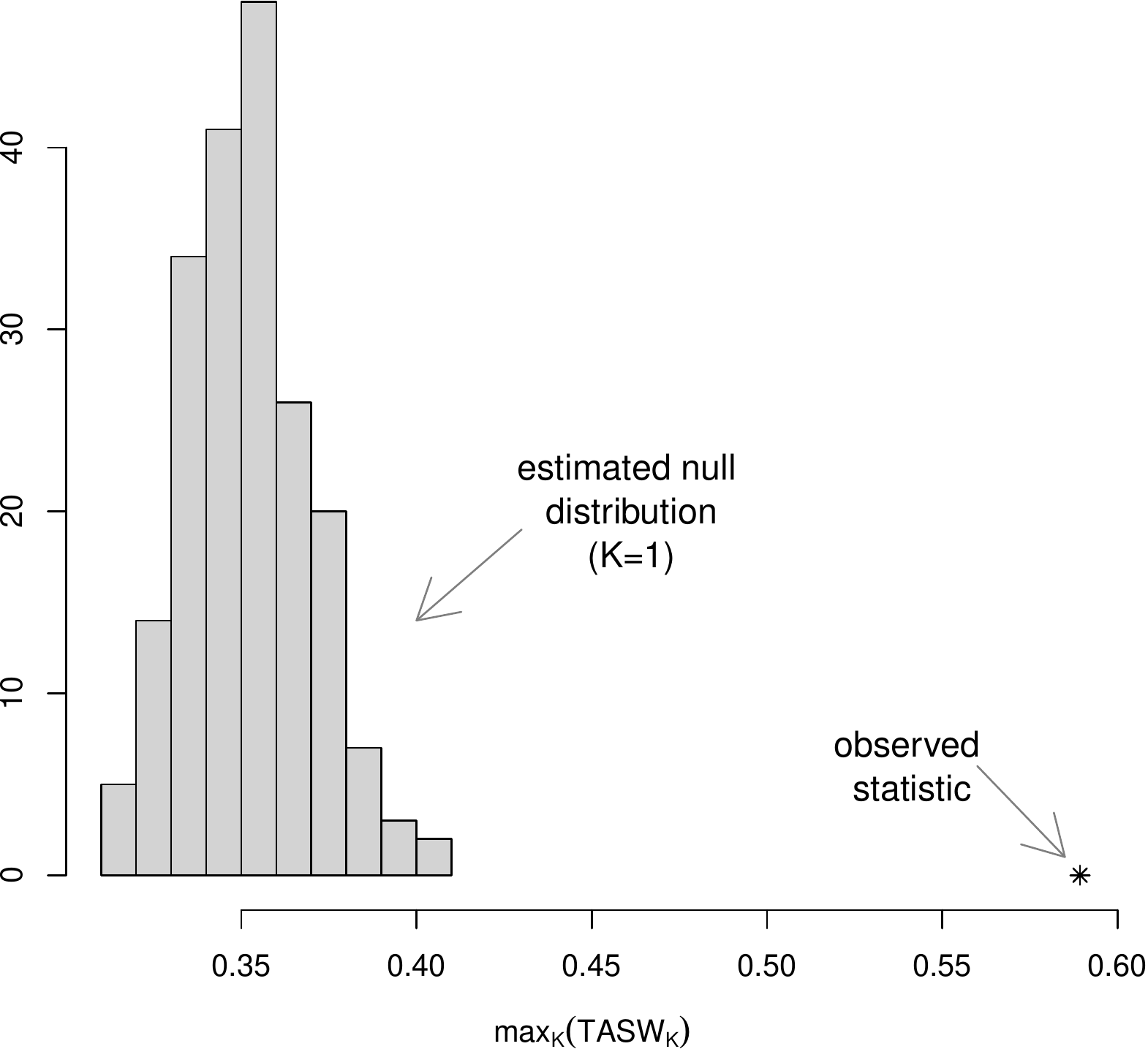}
    \caption{NIR spectra: observed $TASW_{max}$ statistic and an approximation of its null distribution under the hypothesis of no cluster computed using 200 permutations.}
    \label{fig:nirtest}
\end{figure}

\begin{figure}[t]
    \centering
    \includegraphics[width=\textwidth]{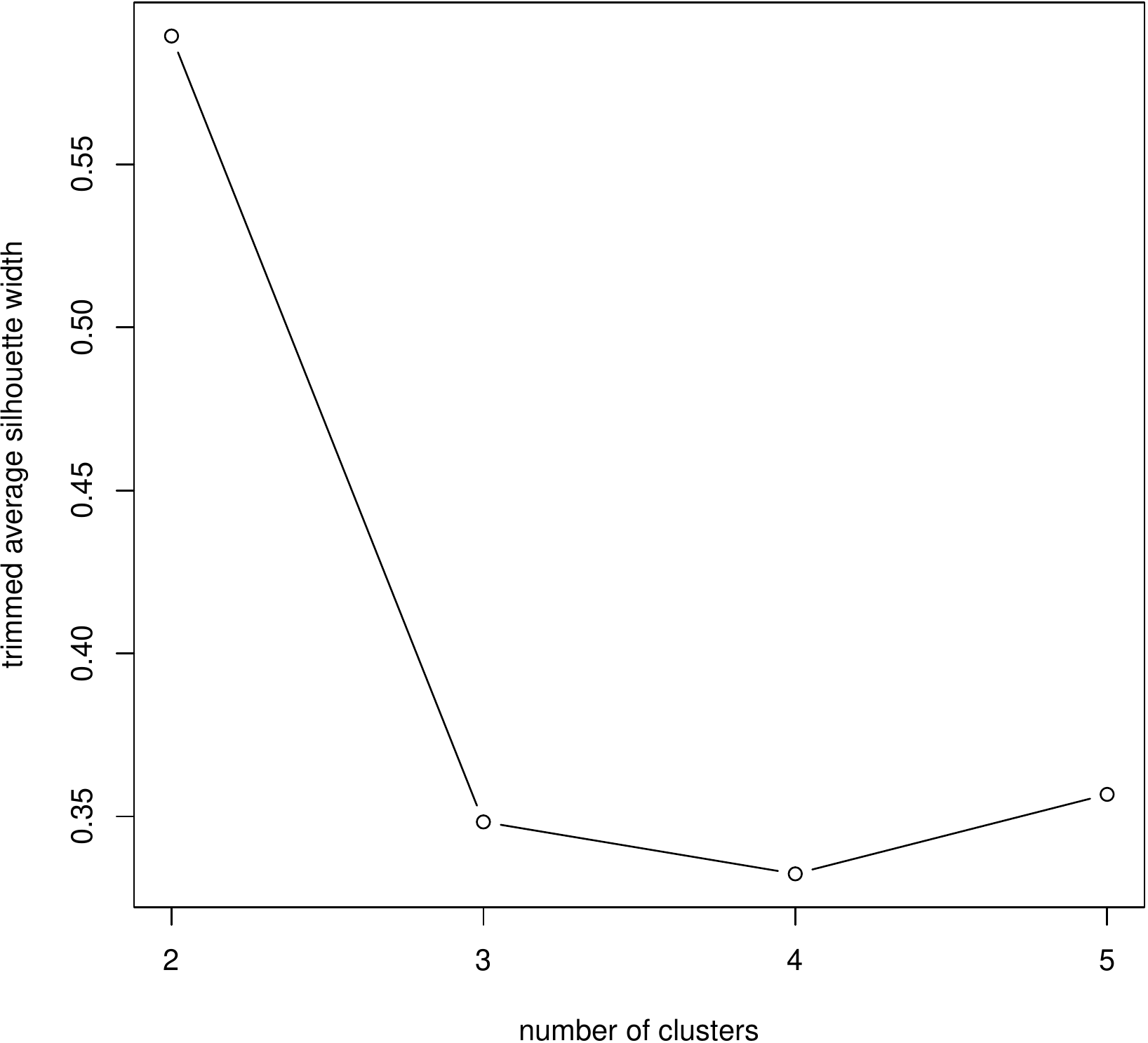}
    \caption{NIR spectra: trimmed average silhouette widths.}
    \label{fig:nirtasw}
\end{figure}

\begin{figure}[t]
    \centering
    \includegraphics[width=\textwidth]{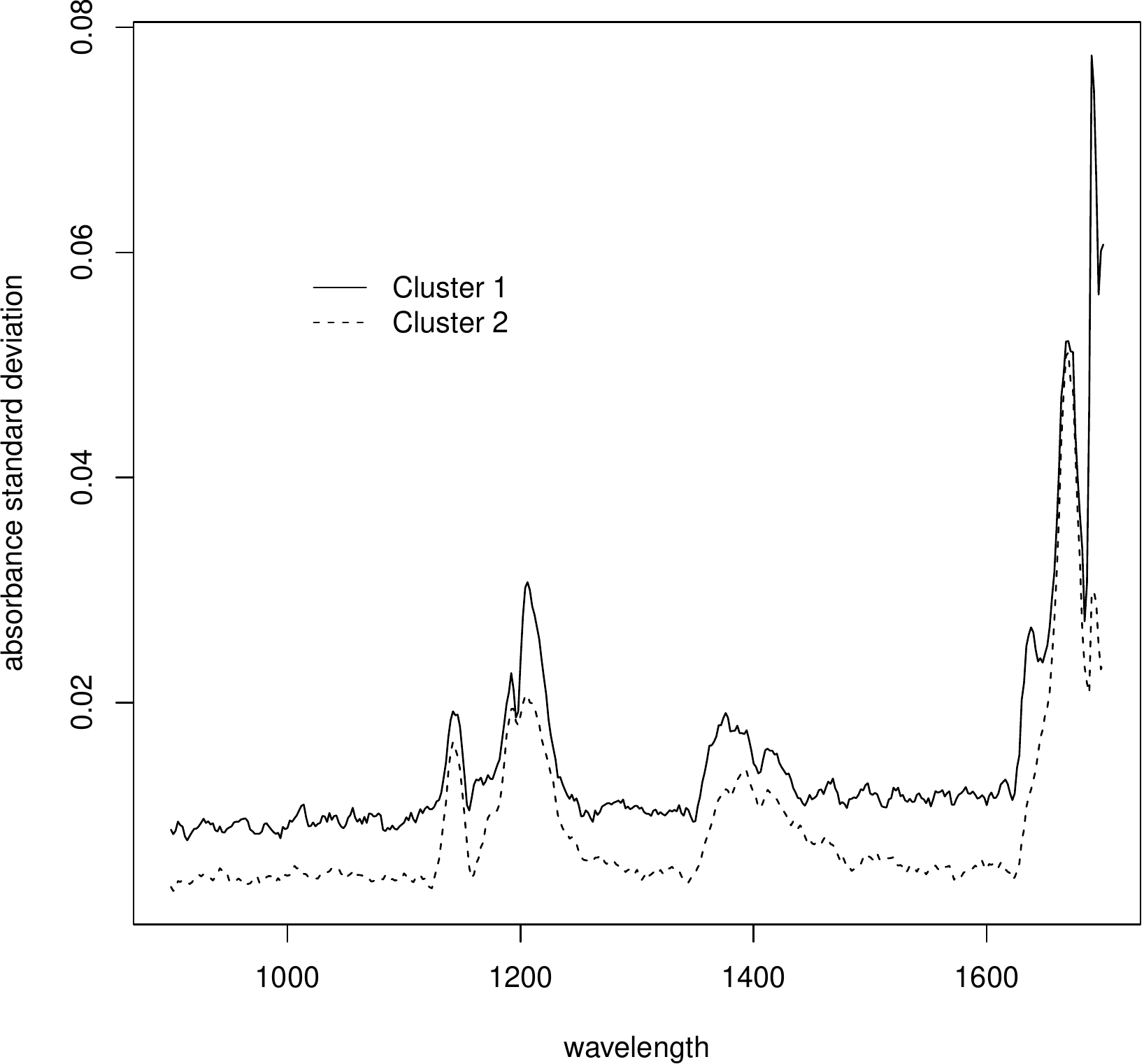}
    \caption{NIR spectra: diagonals of the covariance operators $\overline\Sigma_1$ and 
    $\overline\Sigma_2$.}
    \label{fig:nirsd}
\end{figure}

We applied our clustering algorithm (with $K=2,\ldots,5$) to the $47$ sample covariances estimated using the observations collected on the different days. Figure~\ref{fig:nirtest} shows the observed value of $TASW_{max}$
with an estimate of the null distribution under the hypothesis of no cluster obtained by $200$ random permutations of spectra centered using their day means. The results shown in the figure strongly suggest the presence of some clusters, that is, of some sort of day-to-day variation in the covariance structure. In particular, the $TASW$ profile, displayed in Figure~\ref{fig:nirtasw}, points to two clusters, and the corresponding weights $\pi_{i,1}$ and $\pi_{i,2}$ in Figure~\ref{fig:nirtasw} single out the days from 32 to 37 from the others.
An analysis of the two cluster barycenters $\overline\Sigma_1$ and $\overline\Sigma_2$ reveals differences in the absorbance standard deviations, essentially for each wavelength (see Figure~\ref{fig:nirsd}). 
Thus, the use of the proposed approach allowed us to reach the conclusions reached by \citet{capizzimasarotto_2018} and confirmed by the plant managers about the presence of a transitory instability period.

\begin{figure}[t]
    \centering
    \includegraphics[width=\textwidth]{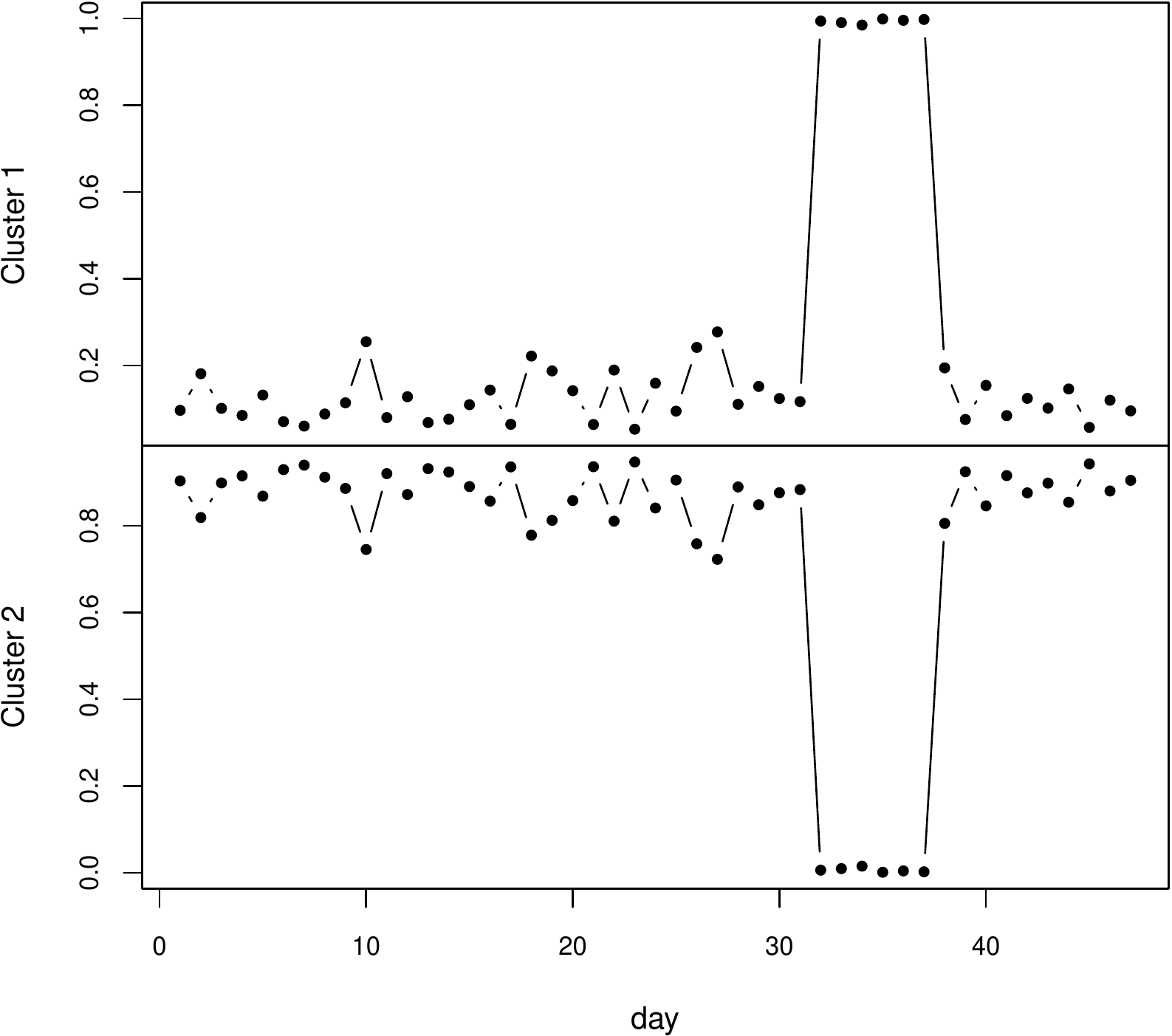}
    \caption{NIR spectra: weights $\pi_{i,j}$ of the solution obtained for $K=2$.}
    \label{fig:nirw}
\end{figure}

\subsection{Phoneme data}
\label{subsec:phoneme}

In this subsection, we consider the dataset available at \url{http://hastie.su.domains/ElemStatLearn/datasets},  
which consist of 4509 log-periodograms of length 256 computed from 
continuous speech frames of male speakers. 
The log-periodograms represent the pronunciation of the five phonemes: 
``aa'',   ``ao'', ``dcl'', ``iy'',  and ``sh''.
With the aim of testing our clustering algorithm, we proceed as follows:
\begin{enumerate}
    \item 
    We randomly select, without replacement,  $30$ subsamples each of size $40$ for each phoneme 
    (thus, obtaining a total of $150$ subsamples).  
    \item
    We apply the proposed cluster algorithm to the corresponding $150$ sample covariances using $K=2,\ldots,10$ (and, of course, ignoring the known phoneme 
    memberships).
\end{enumerate}
To understand whether results were reproducible, the exercise was repeated four times. The code necessary to reproduce part of the first experiment is available in the supplementary material.

\begin{figure}[ht]
    \centering
     \begin{tabular}{cc}
    First experiment & Second experiment\\
    \includegraphics[width=0.5\textwidth]{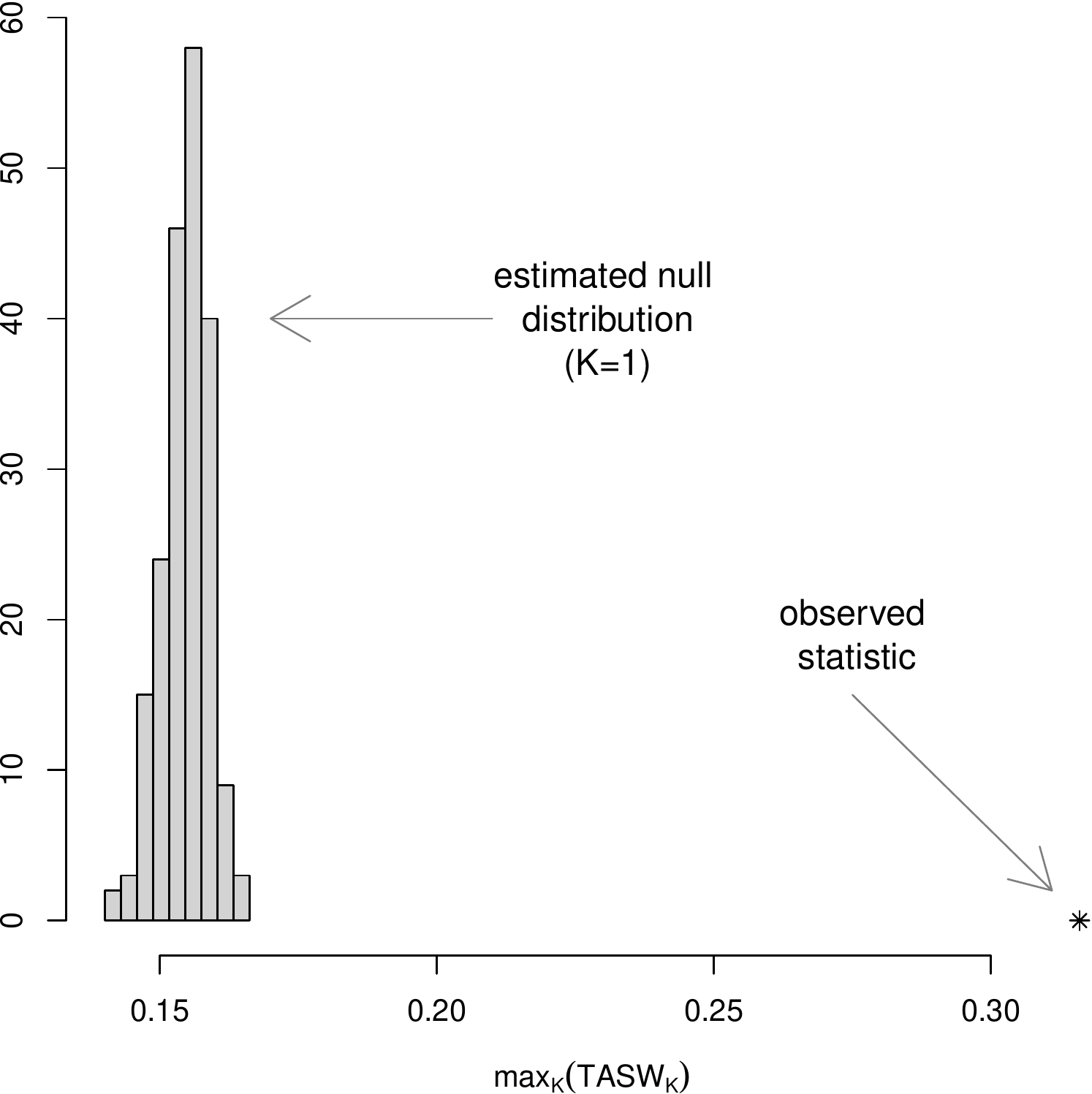} &
    \includegraphics[width=0.5\textwidth]{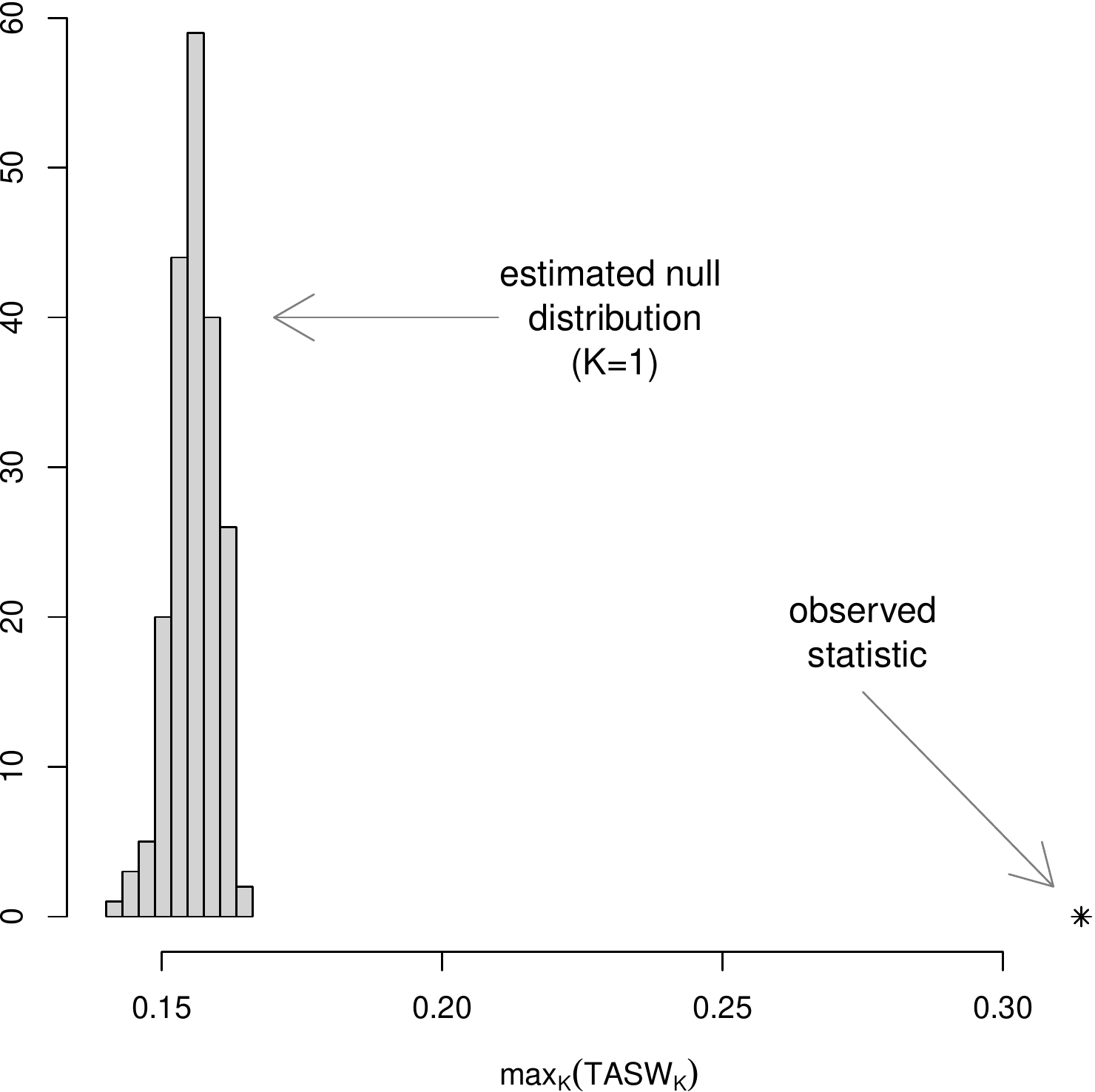} \\
    Third experiment & Fourth experiment \\
    \includegraphics[width=0.5\textwidth]{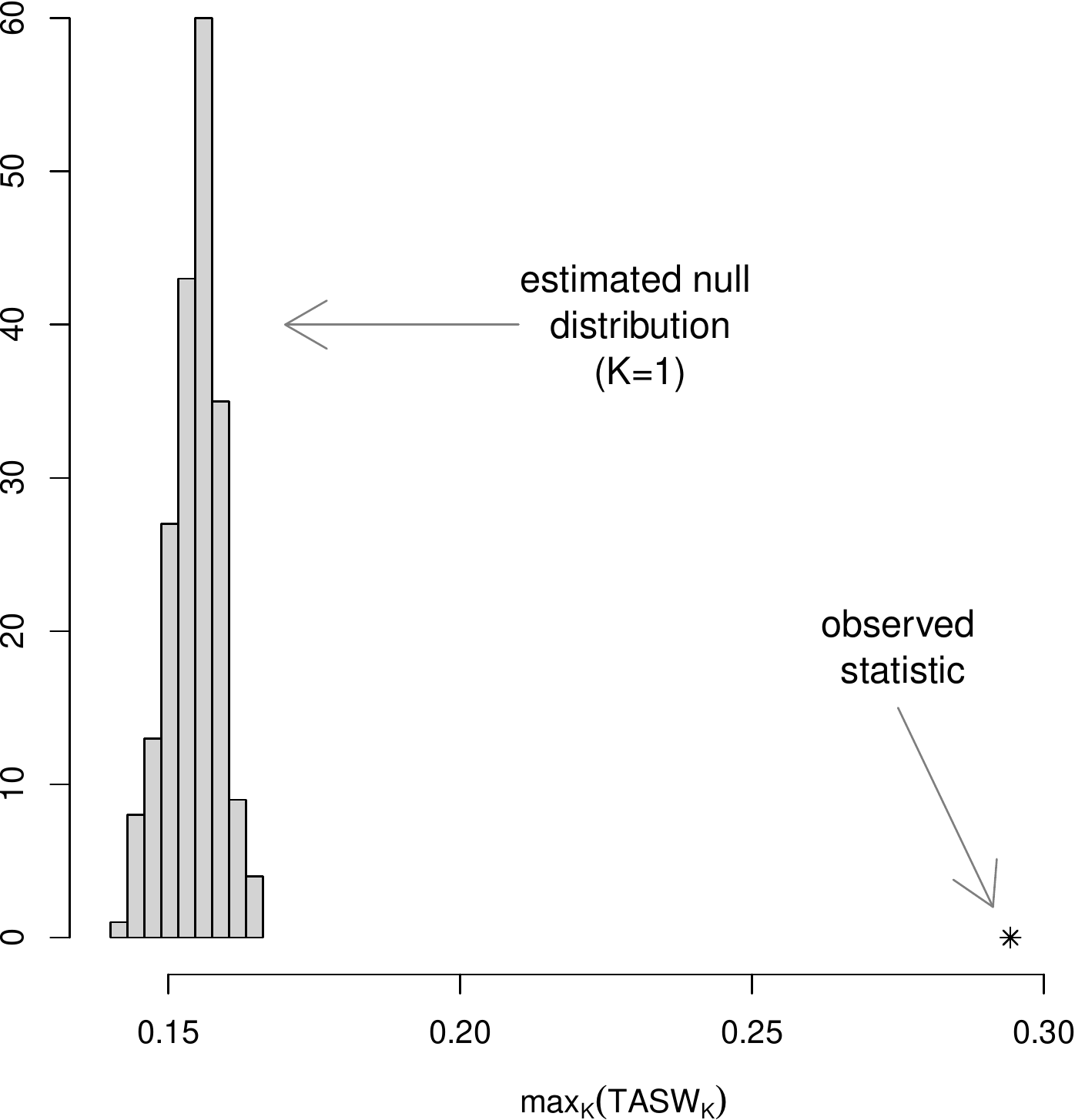} &
    \includegraphics[width=0.5\textwidth]{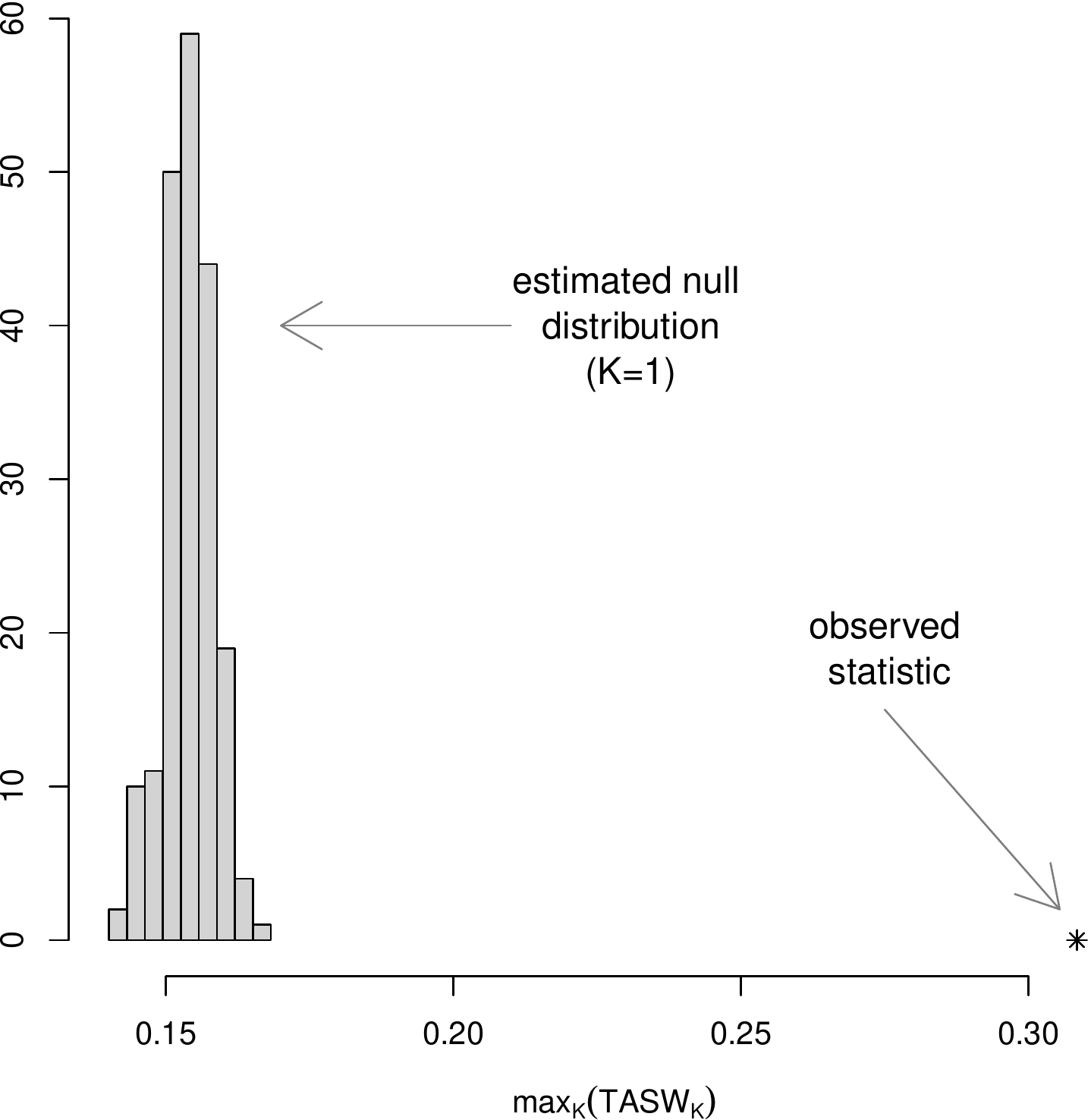} \\
    \end{tabular}
    \caption{Phoneme data: observed $TASW_{max}$ statistics and an approximation of its null distribution under the hypothesis of no cluster computed using 200 permutations.}
    \label{fig:phonemetest}
\end{figure}

\begin{figure}[ht]
    \centering
    \begin{tabular}{cc}
    First experiment & Second experiment\\
    \includegraphics[width=0.5\textwidth]{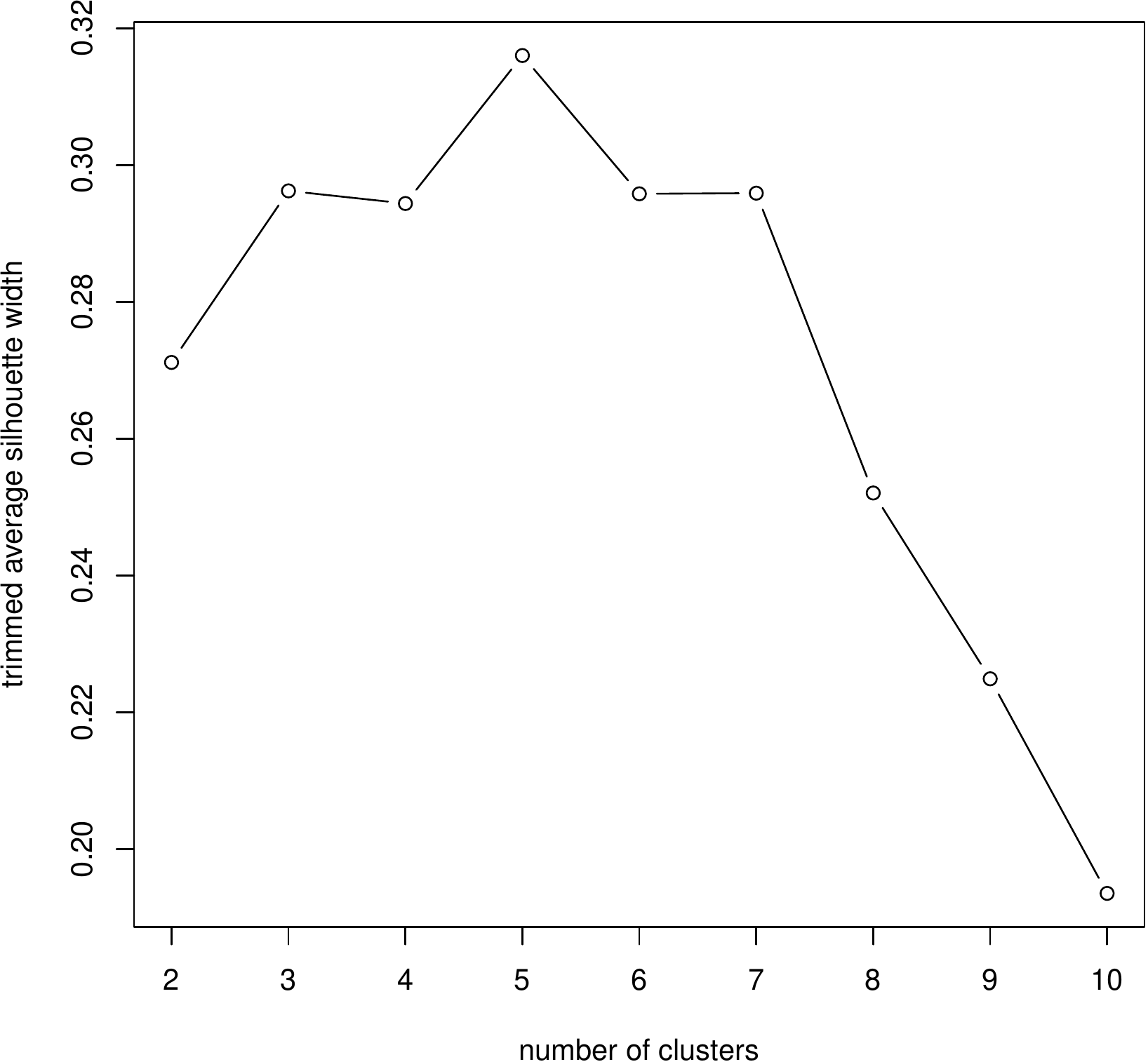} &
    \includegraphics[width=0.5\textwidth]{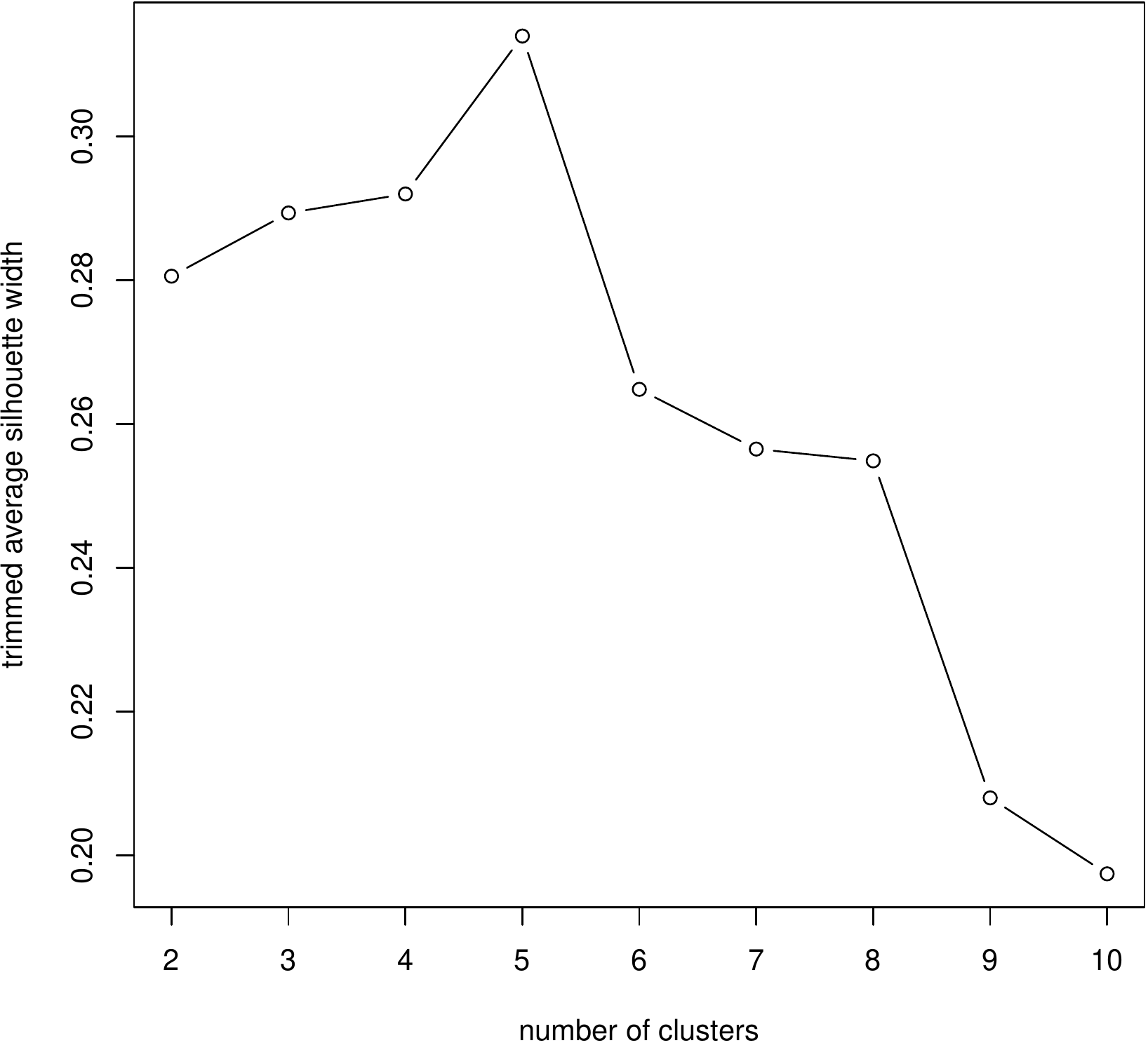} \\
    Third experiment & Fourth experiment \\
    \includegraphics[width=0.5\textwidth]{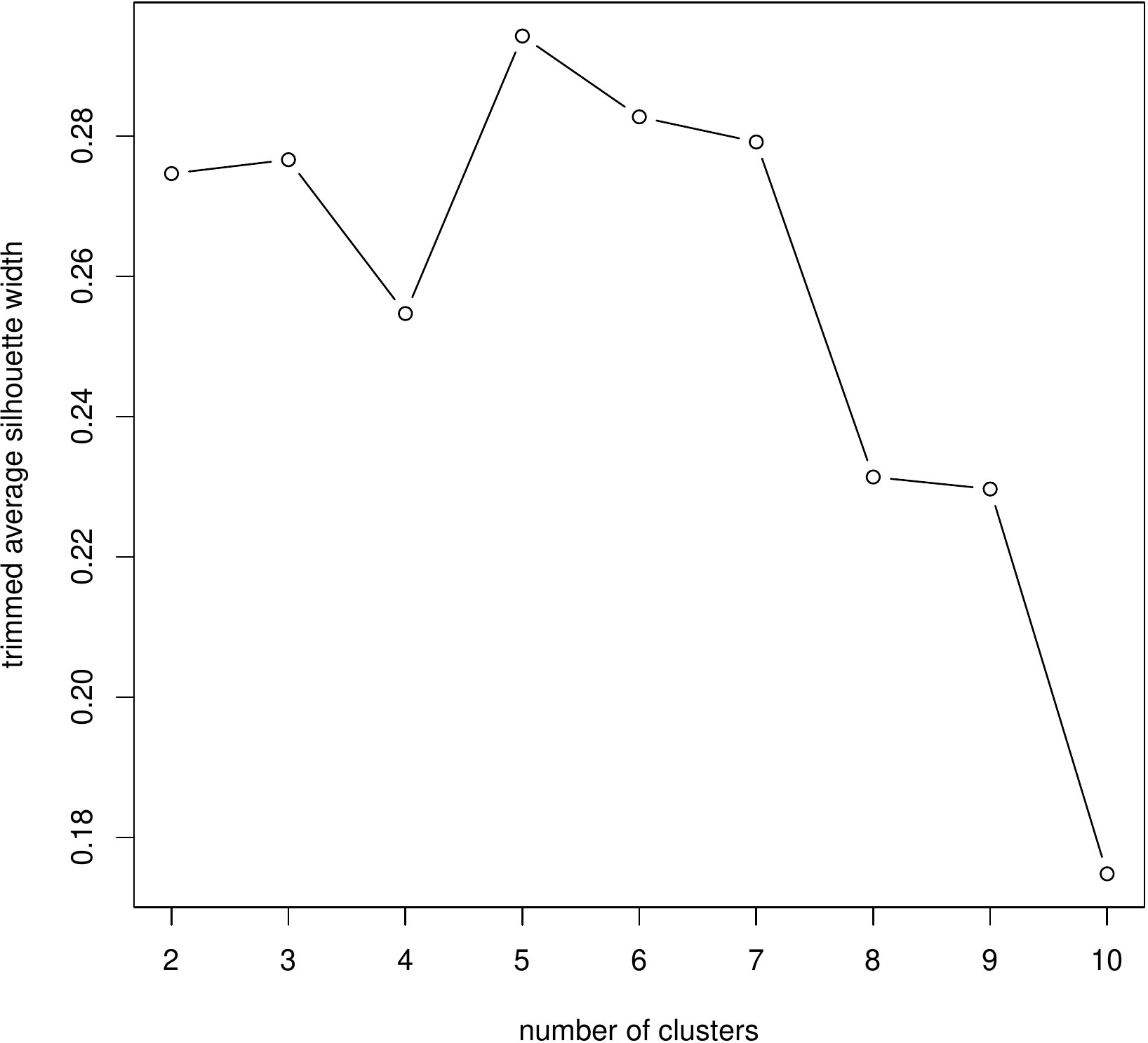} &
    \includegraphics[width=0.5\textwidth]{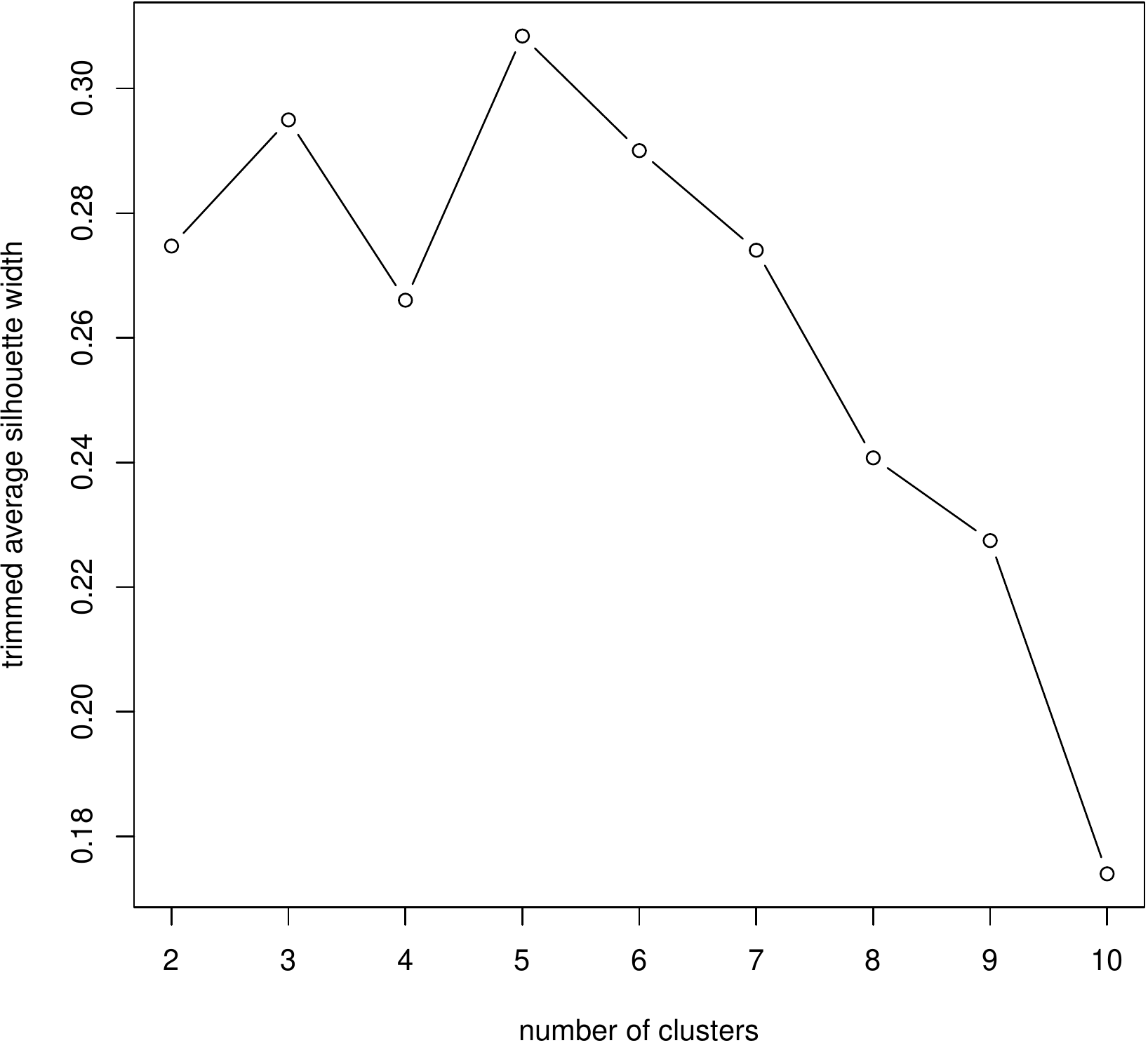} \\
    \end{tabular}
    \caption{Phoneme data: trimmed average silhouette widths.}
    \label{fig:phonemetasw}
\end{figure}

Figure~\ref{fig:phonemetest} shows the observed values of $TASW_{max}$ in the four replications,
with an estimate of the null distribution under the hypothesis of no cluster obtained  
using $200$ random permutations of the (centered) log-periodograms selected during each exercise.
As in the previous subsection, results strongly suggest the presence of a cluster structure in the data. The $TASW$ profiles, displayed in Figure~\ref{fig:phonemetasw}, suggest $K=5$ in each of the four experiments.  The visual representation, obtained by multidimensional scaling, of the
\begin{itemize}
    \item[(i)] ``true'' covariances of the five phonemes, that is, the covariances estimated using all the data for each phoneme, and
    \item[(ii)] the five cluster barycenters $\overline\Sigma_1, \ldots, \overline\Sigma_5$
\end{itemize}
shows that the algorithm was consistently able to reconstruct the presence of the five phonemes.

\begin{figure}[ht]
    \centering
    \begin{tabular}{cc}
    First experiment & Second experiment\\
    \includegraphics[width=0.5\textwidth]{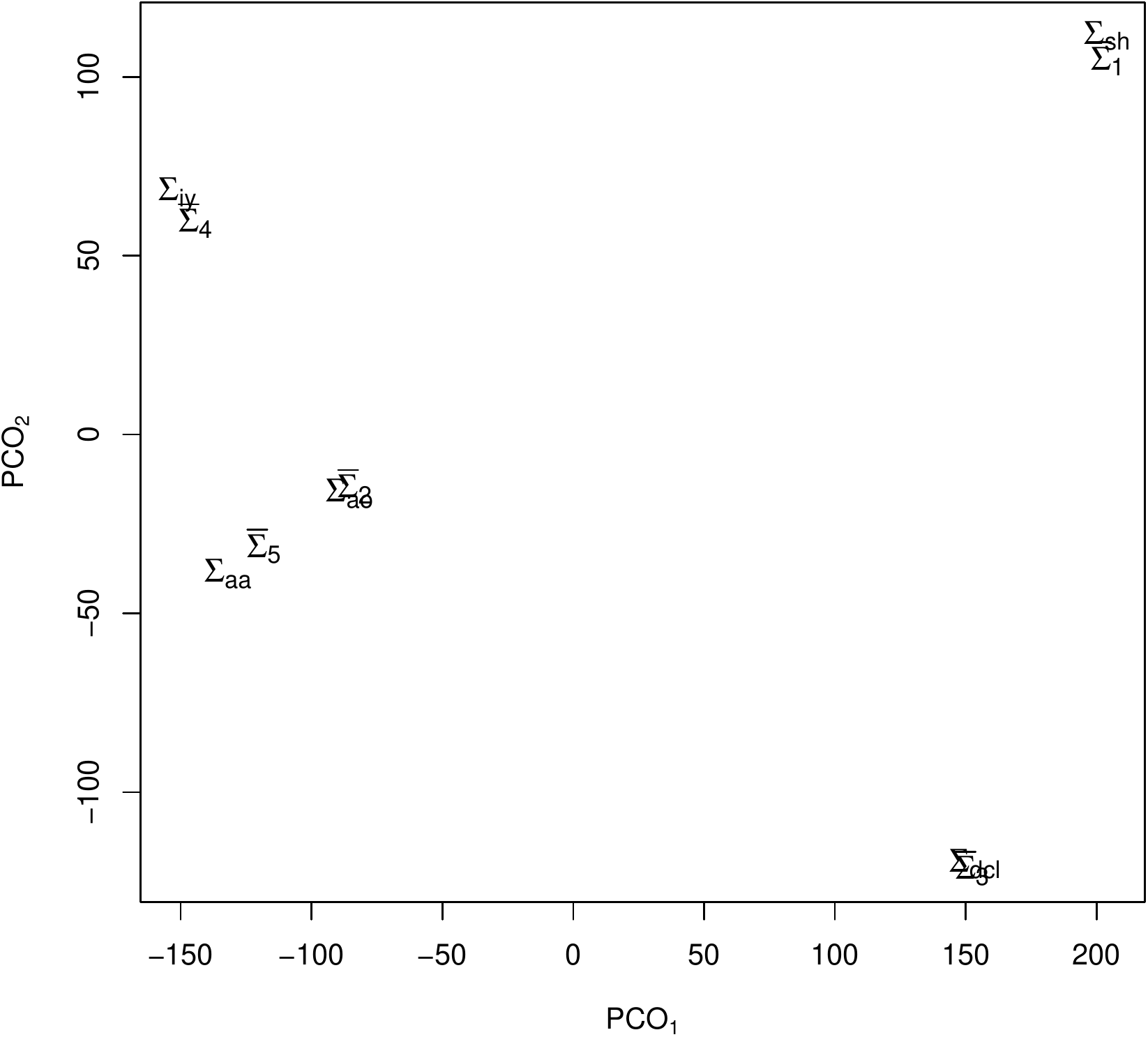} &
    \includegraphics[width=0.5\textwidth]{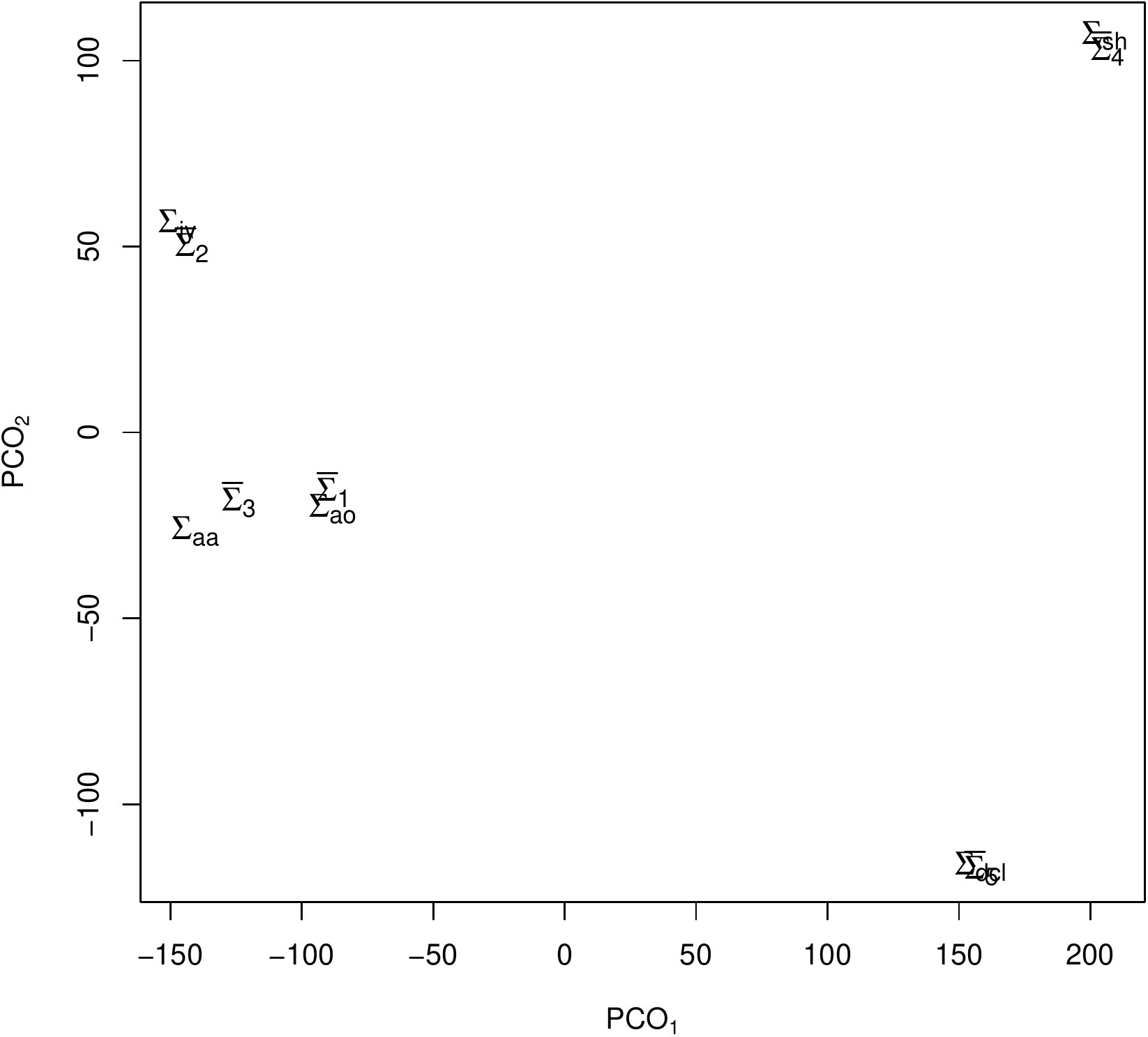} \\
    Third experiment & Fourth experiment \\
    \includegraphics[width=0.5\textwidth]{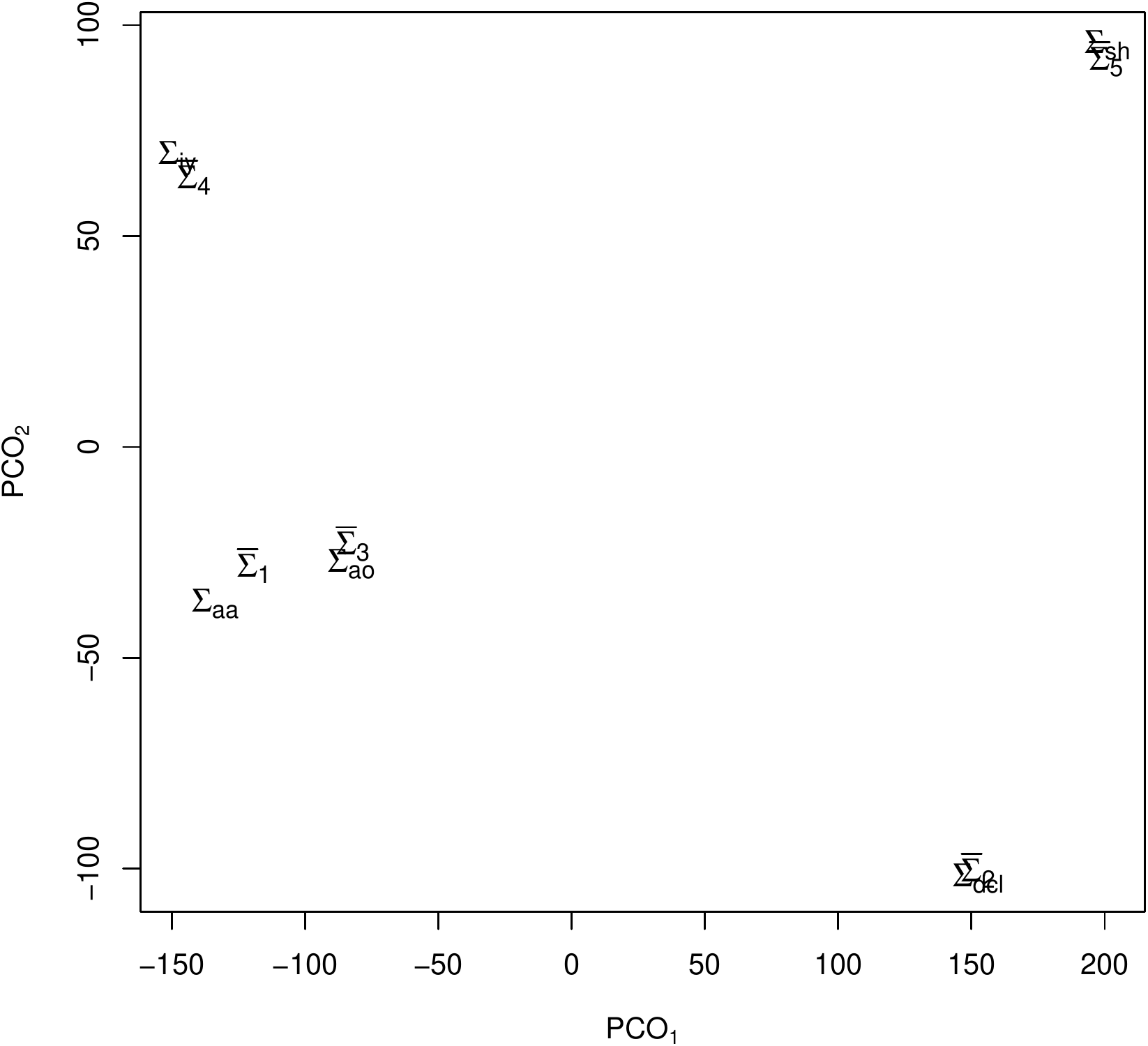} &
    \includegraphics[width=0.5\textwidth]{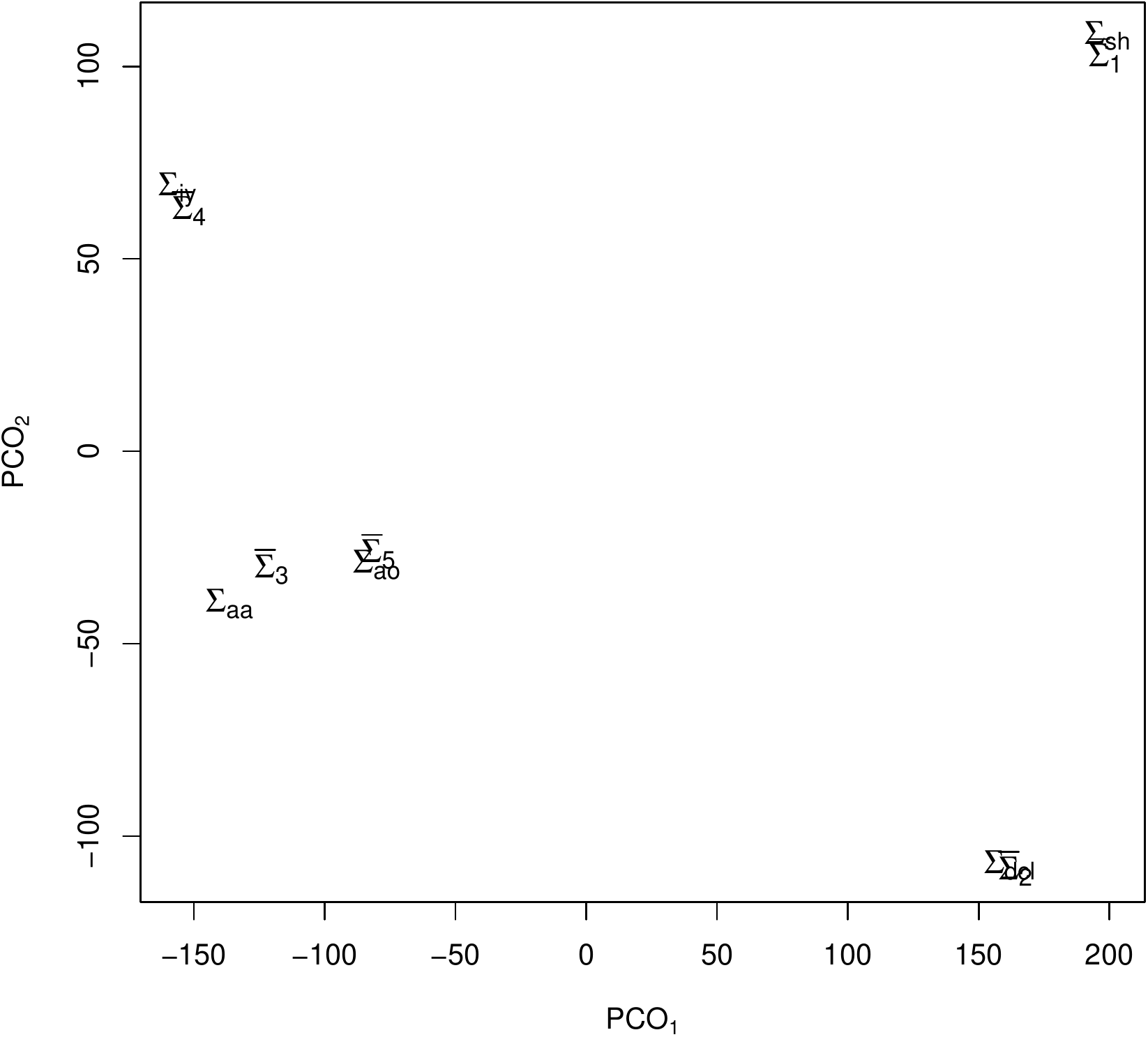} \\
    \end{tabular}
    \caption{Phoneme data: multidimensional scaling representation 
    (principal coordinates analysis) based on the Wasserstein-Procrustes distance
    of the $5$ ``true'' phoneme covariances 
    ($\Sigma_{\text{pnoneme}}$ with phoneme=``aa'', ``ao'', ``dcl'', ``iy'' and ``sh''), and of the  estimated cluster barycenters ($\overline\Sigma_j$ with $j=1,\ldots,5$).}
    \label{fig:phonememds}
\end{figure}

It is also interesting to observe that the 
partition matrix provides an idea of the degree of overlap between groups. For example, the  cross-product of the matrix obtained for $K=5$ 
during the first experiment is
\[
(P^5)'P^5=
\begin{pmatrix}
29.1  \\
0.1  & 22.5 \\
0.9  &  0.2 & 28.2  \\
0.1  &  1.8 &0.1 & 25.0\\
0.1  &  5.3  & 0.2 & 2.7 & 22.3\\
\end{pmatrix}.
\]
A perusal of this matrix reveals that
\begin{itemize}
    \item the first and third identified clusters are essentially isolated; indeed, the corresponding 
    off-diagonal entries are quite small;
    \item the second and fifth clusters have a certain degree of overlap  (the $(5,2)$ entry is the largest between the off-diagonal ones); there also exists a more limited overlap between these two clusters and the fourth cluster. 
\end{itemize}
A look at the graph at the top left in Figure~\ref{fig:phonememds} makes clear the reasons for what we have just described: the second and fifth clusters correspond to the phonemes ``ao'' and ``aa'', respectively. These two phonemes are, as expected, relatively hard to distinguish. The fourth cluster corresponds to the ``iy'' phoneme,  which is closer to ``ao'' and ``aa'' than to the other two phonemes. Finally, the first and third clusters correspond to the phonemes ``sh'' and ``dcl'',  
whose pronunciation is easier to distinguish.  
Analogous results were obtained for the other three replication of the exercise.

\section{Conclusions}
\label{sec:conclusion}
We introduced a novel way to explore similarities among a group of functional covariance operators via a soft clustering mechanism  based directly on the Wasserstein-Procrustes distance. The proposed framework is particularly useful in situations where there is no clean-cut cluster separation, and a certain degree of cluster overlapping could be envisaged. An attractive feature of the algorithm is that it allows to adjust for the expected degree of such overlapping, through the modulation of the average entropy level.  
Moreover, we contributed a novel cluster quality index, which we called the 
trimmed average silhouette width (TASW), which adapts the fast silhouette width considered by \citet{van_der_laan_new_2003} to the covariance setting and ``trims'' it in a way that only clusters that are credibly far away (more than a given threshold) are considered. 
The TASW is shown empirically to behave well in simulations, and correctly estimate the proper number of clusters; however, different thresholds could be employed in the definition. We leave the details about the behaviour of different thresholds to future work.  We also illustrate the use of the proposed methodology with an application to a gasoline near-infrared spectra dataset from \citet{capizzimasarotto_2018}, and showed that the suggested approach allows us to reach the same conclusions as \citet{capizzimasarotto_2018} in a neat and simple way. 

As a conclusive remark about future research, we would like to mention leveraging  the geometric interpretation of the Wasserstein-Procrustes distance given in \citet{masarotto2018procrustes} that allows for the definition of other clustering algorithms, 
which perform partitioning on the tangent space of the nonlinear manifold inhabited by the covariance operators. Lifting the covariances to the tangent space
makes it possible, and indeed interpretable, to shift the non-linear problem to a linear space, and for example, carry on traditional $k$-means based on the Hilbert--Schmidt distance. This intuition can be expanded further to develop  hard and soft clustering methods.

\appendix
\section{Proof of the proposition of Subsection \ref{subsec:comp_impl}}

Point 1 of the proposition follows directly from \eqref{eqn:obj}. 
As for 2, observe that \eqref{eqn:poptimal}--\eqref{eqn:etaconstraint} can be easily obtained by introducing the Lagrange function
  \[
  {\cal L}=  \sum_{i=1}^N\sum_{j=1}^K \pi_{i,j} d_{i,j}
  +\eta\left(\sum_{i=1}^N\sum_{j=1}^K \pi_{i,j}\log(\pi_{i,j})+N E\right)
  + \sum_{i=1}^N \lambda_i\left(\sum_{j=1}^K \pi_{i,j}-1\right),
  \]
  where $d_{i,j}=\bigl(n_i-1\bigr)\Pi\bigl(\widehat\Sigma_i, \overline\Sigma_j\bigr)$.
In particular, equation \eqref{eqn:pformula} for the weights follows directly from 
the first-order conditions for $\lambda_i$ and $\pi_{ij}$ 
  \begin{align*}
  \dfrac{\partial{\cal L}}{\partial\pi_{i,j}}&=d_{i,j}+\eta\bigl(\log(\pi_{i,j}\bigr)+1)+\lambda_i=0,
  &(i=1,\ldots,N; j=1,\ldots,K),\\
  \dfrac{\partial{\cal L}}{\partial\lambda_{i}}&=\sum_{j=1}^K \pi_{i,j}-1=0,
  &(i=1,\ldots,N).
  \end{align*}
Therefore, the constrained optimization problem becomes 
\[
\min_{\eta} \Phi(\eta) \text{ subject to } \Psi(\eta) = NE,
\]
where $\Phi(\eta) = \sum_{i,j} \pi_{i,j}(\eta) d_{i,j}$ and
$\Psi(\eta)=-\sum_{i,j} \pi_{i,j}(\eta)\log \pi_{i,j}(\eta)$. For simplicity, suppose that all the distances 
$d_{i,j}$ are distinct and greater than zero. 
This assumption will be true with probability one in applications (at least when the $X_{i,j}$'s are real). 
However, in any case, it is easy to adapt the following proof to the general case.
Under the previous assumption, it is easy to show that
\begin{enumerate}
    \item[(a)] 
    for every $i$ and $j$
    \[
    \lim_{\eta \rightarrow -\infty} \pi_{i,j}(\eta) = \lim_{\eta \rightarrow +\infty} \pi_{i,j}(\eta)=\dfrac{1}{K}.
    \]
    As a consequence, 
    \begin{align*}
    &\lim_{\eta \rightarrow -\infty} \Phi(\eta) = \lim_{\eta \rightarrow +\infty} \Phi(\eta) = \Phi_\infty\\
    &\lim_{\eta \rightarrow -\infty} \Psi(\eta) = \lim_{\eta \rightarrow +\infty} \Psi(\eta) = N\log K
    \end{align*}
    where
    \[
    \Phi_\infty=\dfrac{1}{K}\sum_{i,j} d_{i,j}.
    \]
    \item[(b)] 
    for every $i$
    \begin{align*}
    \lim_{\eta \rightarrow 0^-} \pi_{i,j}(\eta) &=
    \begin{cases}
      0 & \text{if } j \ne \arg \max_r d_{i,r} \\
      1 & \text{if } j = \arg \max_r d_{i,r} \\
    \end{cases}\\
    \intertext{and}\\
    \lim_{\eta \rightarrow 0^+} \pi_{i,j}(\eta) &=
    \begin{cases}
      0 & \text{if } j \ne \arg \min_r d_{i,r} \\
      1 & \text{if } j = \arg \min_r d_{i,r} \\
    \end{cases}.
    \end{align*}
    Thus, 
    \[
    \lim_{\eta \rightarrow 0^-} \Phi(\eta) = \Phi_{0^-} 
    \text{ and } 
    \lim_{\eta \rightarrow 0^+} = \Phi_{0^+} 
    \]  
   with
   \[
    \Phi_{0^-} = \sum_i \max_j d_{i,j} \\ 
    \text{ and }
    \Phi_{0^+} = \sum_i \min_j d_{i,j}.
    \]
    In addition,
    \[ \lim_{\eta \rightarrow 0} \Psi(\eta) = 0.\]
    \item[(c)] 
    for every $\eta \ne 0$,
    \[
    \dfrac{d\Phi(\eta)}{d\eta} = \dfrac{V^2(\eta)}{\eta^2}
    \text{ and }
    \dfrac{d\Psi(\eta)}{d\eta} = \dfrac{V^2(\eta)}{\eta^3}
    \]
    where, defined $\overline d_i=\sum_J \pi_{i,j}d_{i,j}$,
    \[
    V^2(\eta) = \sum_{i,j} \pi_{i,j}\bigl(d_{i,j}-\overline d_{i}\bigr)^2 .
    \]
\end{enumerate}
When $0 < E < \log K$, the previous properties show that
$\Psi(\eta)$ monotonically decreases from $N\log K$ to $0$ when $\eta$ goes from $-\infty$ to $0$, and monotonically increases from $0$ to $N\log K$ when $\eta$
goes from $0$ to $+\infty$. 
Thus, the equation $\Psi(\eta)=NE$ has exactly two roots $-\infty < \eta^- < 0 < \eta^+ < +\infty$.
However, (a)--(c) also guarantee that $\Phi(\eta^+) < \Phi_\infty <   \Phi(\eta^-)$, and, therefore, we only need  to compute the positive root $\eta^+$. On the other hand, if $E=0$ (or $\log K$), the desired solution is given by the second degenerate distribution given in (b) (or by the uniform distribution given in (a)). 

\begin{supplement}
\stitle{}
\sdescription{
\vspace*{-3\baselineskip}
\begin{description}
\item[wasserstein.R:] \texttt{R} implementation of the proposed clustering method; the functions are documented inside the file.
\item[phonemes.R:] \texttt{R} script that reproduces part of the phoneme example presented in Subsection \ref{subsec:phoneme}. 
\end{description}
}
\end{supplement}

\bibliographystyle{imsart-nameyear}
\bibliography{fda}

\end{document}